\documentclass[preprint,11pt]{elsarticle}
\usepackage[letterpaper,top=1in, bottom=1in, left=1in, right=1in,marginparwidth=1in]{geometry}
\usepackage{xcolor}
\usepackage{graphicx}
\usepackage{amsmath,amsfonts,amssymb,amsthm,epsfig,epstopdf,url,array}
\usepackage{amsfonts}
\usepackage{float}
\usepackage{booktabs}  
\usepackage[english]{babel}
\usepackage[autostyle]{csquotes}
\usepackage{subfig}
\theoremstyle{plain}
\usepackage{multirow}
\usepackage[final]{changes}
\usepackage[switch]{lineno}

\begin{document}

\begin{frontmatter}
\title{Understanding charging dynamics of fully-electrified taxi services using large-scale trajectory data}

\author[SZTU_address]{Tian Lei}
\ead{leitian@sztu.edu.cn}

\author[ua_address]{Shuocheng Guo}
\ead{sguo18@ua.edu}

\author[ua_address]{Xinwu Qian\corref{cor}}
\cortext[cor]{Corresponding author.}
\ead{xinwu.qian@ua.edu}

\author[SZTU_address]{Lei Gong}
\ead{gonglei@sztu.edu.cn}

\address[SZTU_address]{College of Urban Transportation and Logistics, Shenzhen Technology University, Shenzhen, China, 518118}

\address[ua_address]{Department of Civil, Construction and Environmental Engineering, The University of Alabama, Tuscaloosa, AL, United States, 35487}

\begin{abstract}
An accurate understanding of \enquote{when, where and why} of charging activities is crucial for the optimal planning and operation of E-shared mobility services. In this study, we leverage a unique trajectory of a city-wide fully electrified taxi fleet in Shenzhen, China, and we present one of the first studies to investigate charging behavioral dynamics of a fully electrified shared mobility system from both system-level and individual driver perspectives. The electric taxi (ET) trajectory data contain detailed travel information of over 20,000 ETs over one month period. By combing the trajectory and charging infrastructure data, we reveal remarkable regularities in infrastructure utilization, temporal and spatial charging dynamics as well as individual driver level charging preferences. Specifically, we report that both temporal and spatial distributions of system-level charging activities present strong within-day and daily regularities, and most charging activities are induced from drivers' shift schedules. Further, with 425 charging stations, we observe that the drivers show strong preferences over a small subset of charging stations, and the power-law distribution can well characterize the charging frequency at each charging station. Finally, we show that drivers' shift schedules also dominate the individual charging behavior, and there are strikingly stable charging patterns at the individual level. The results and findings of our study represent lessons and insights that may be carried over to the planning and operation of E-shared mobility in other cities and deliver important justifications for future studies on the modeling of E-shared mobility services. 
\end{abstract}


\end{frontmatter}

\section{Introduction}

The growth of electric vehicles (EV) in replacing traditional gasoline vehicles has brought significant changes to urban mobility's composition and behavioral characteristics. As an essential part of the urban transportation system, shared mobility such as taxis and for-hire vehicles in multiple metropolitan areas have shifted to utilizing EVs with policy incentives initiated by local governments~\cite{sathaye2014optimal}. Several big cities, such as New York, Shenzhen, and Beijing, have launched initiatives to promote the use of EVs in the taxi industry~\cite{wei2020real}, leading to a rapid growth in the adoption of EVs in urban shared mobility systems. For instance, from 2016 to the end of 2020, Shenzhen's buses and taxis have all been fully electrified~\cite{report}. The increasing penetration of EVs will further influence the behavior of shared mobility considering its difference of charging needs when compared to the refueling of traditional gasoline vehicles~\cite{yang2016predicting}. The coordination among charging demand, charging facility distribution, and charging time and duration~\cite{yang2018design,kempton2016electric} are becoming essential factors in the operation of electrified mobility services. As a result, there is a pressing need for an accurate understanding of the charging dynamics for electrified shared mobility services in order to assist the optimal operation and the planning of urban transportation facilities.

While problems such as electric taxi (ET) dispatching\cite{tseng2018improving,song2019demand} and charging scheduling~\cite{tian2016real,yang2018distributed, you2020charging,wei2020real,shen2020integrated} are of significant interest for the electrified shared mobility (E-shared mobility), real-world evidence from large-scale E-shared mobility systems is still very limited. Some recent studies attempted to address this issue by analyzing the charging and operating behavior of a small-scale ET fleet. However, compared to the giant scale of urban shared mobility services, those case studies can only cover a very small subset of vehicle fleet~\cite{rao2018modeling,fraile2018using,tian2014understanding}. In addition, due to the lack of real-world evidence, early studies on EV modeling attempted to make various assumptions on the charging behavior, such as competition among EV users ~\cite{he2013optimal,chen2020optimal,liu2017locating}, drivers' preferences over charging stations~\cite{ke2019modelling,song2019demand} and drivers' rational behavior in trading off among charging price, future income, and the time of charging~\cite{masoum2016hybrid}. We note that none of the studies justified their assumptions with practical evidence, which act as decisive factors that affect outcomes and conclusions. As a consequence, the validity of the reported findings and insights remain questionable and would require further justification. This motivates us to deliver systematic and statistically meaningful insights on the charging and operating behavior under a fully E-shared mobility system.

In light of the existing gaps in the literature, this study presents one of the first empirical evaluations on the charging dynamics of a large-scale fully electrified shared mobility service: the ET system in Shenzhen, China. It should be noted that Shenzhen is one of the leading cities globally in charging infrastructure development, where over 440,000 EVs (12.4\% of all vehicles) are running on the road as of September 2020~\cite{Shenzhen}. Therefore, the large-scale ET data makes our case study unique and representative in the era of transportation electrification. More specifically, this study concerns the system and individual charging dynamics of over 20,000 ETs and 38,000 drivers, which translates to nearly 30,000 daily charging activities across 425 charging stations. In this study, we first model the temporal and spatial charging dynamics at the system level by mining the charging activities from GPS trajectory data of the large-scale ET fleet. Station-level charging dynamics are further discussed to identify factors that affect drivers' preferences over charging stations. Finally, we leverage \added{a spatiotemporal topic modeling approach} to investigate charging preferences at individual driver level. The combination of the three-level analyses provides a comprehensive understanding on when, where and why the charging activities take place. The results of our study therefore provide significant insights and invaluable evidence in assisting the planning and operation of ET service and E-shared mobility services. \added{In particular, the major contributions of this article include:}
\begin{itemize}
    \item \added{To the best of our knowledge, this is the first study to investigate the interaction of a fully electrified shared mobility system and the large-scale charging infrastructures, which enables in-depth understanding about the charging dynamics of a fully electrified shared mobility system in urban societies and the charging behavior of ET drivers.}
    \item \added{This study examines the validity of the hypotheses applied in previous studies about EV charging behavior and the appropriateness of equilibrium assumption in the EV facility planning problem. The findings will pave the way for future charging behavior modelling of E-shared mobility systems. Through identifying underlying factors that influence ET drivers' charging choices with realistic demonstration, further charging behavior model can be built to quantify the relationship between certain aspects and ET drivers' charging decisions.}
    \item \added{Finally, the present work draws import implications for practice regarding charging infrastructure planning as well as ET charging management. Conclusions about charging dynamics, charging facility usage and drivers' charging choices obtained from station-level analysis as well as individual-level charging behavior model can provide valuable reference for further deployment of charging infrastructures and optimized resource allocation.} 
\end{itemize}

We summarize the organization of our study as follows. Section 2 reviews the literature on related studies. Section 3 introduces the trajectory data, charging station data, and driver shift data used in this study. Section 4 introduces the methods and metrics for modeling and evaluating the ET charging dynamics. Section 5 presents the results regarding the spatial-temporal charging dynamics, station-level charging dynamics, and individual-level charging behavior. Furthermore, in Section 6, we will discuss the charging behavior of the ET drivers and note the potential implications. Finally, conclusions and future directions will be included in Section 7.

\section{Literature Review}

\subsection{EV charging behavior modeling}
The rapid development of EVs has resulted in a sprout of studies in charging demand prediction~\cite{yun2018prediction,bae2011spatial}, charging behavior modeling~\cite{zhao2013electric} and the planning of the charging infrastructure~\cite{brooker2015identification,khaksari2021sizing}. A large body of studies have been conducted on system modeling of E-shared mobility service. Particularly for ETs, the topics \added{include} charging facility planning for ETs~\cite{chen2017data,wang2021taxi}, ET service/charging  strategy~\cite{tseng2018improving,tian2016real,yang2018distributed, you2020charging}, ET fleet dispatching~\cite{song2019demand} and scheduling strategy~\cite{shen2020integrated}. To enable the modeling framework, several key assumptions on charging behavior are made and can be grouped into three categories that are related to system-level charging behavior coupled with charging station usage, drivers' preference for charging stations, and drivers' preference for charging time, respectively. A summary of these studies can be found in Table~\ref{tab:assumption}. 

\begin{table}[H]
\small
\caption{Assumptions about charging behavior in existing studies}
\begin{tabular}{p{4cm}p{4cm}p{3cm}p{4cm}}
\toprule  
Charging behavior & Major assumption  & Source & Problem\\
\hline
Competing behavior among EV users & Route choice and charging time equilibrium & ~\cite{he2013optimal},
~\cite{he2014network},
~\cite{liu2017locating},
~\cite{zheng2017traffic},
~\cite{xu2017network},
~\cite{huang2020electric},
~\cite{chen2020optimal}, ~\cite{chen2016optimal}, ~\cite{chen2017deployment},
~\cite{zhou2020optimal},
~\cite{sun2020integrated} & Charging facility planning\\
\hline
\multirow{2}{4cm}{Drivers' preference for charging station location} & Always go to the nearest charging station & ~\cite{jia2012optimal}, ~\cite{tu2016optimizing}, ~\cite{ke2019modelling}, ~\cite{song2019demand}, ~\cite{hu2018analyzing} & ET charging schedule and dispatching strategy; EV feasibility \\
\cline{2-4} ~ & Drivers tend to charge at preferred charging station & ~\cite{ahn2016analysing}, ~\cite{dong2017rec} & ET charging schedule \\
\hline
Drivers' choice for time of charging & Drivers prefer overnight charging & 
~\cite{richardson2011optimal}
~\cite{masoum2016hybrid}, ~\cite{heymann2017mapping}, ~\cite{zhou2020optimal}, ~\cite{lajunen2018lifecycle}, ~\cite{houbbadi2019optimal}, ~\cite{olsson2017lessons}, ~\cite{moniot2019feasibility} & EV charging coordination and EV battery range optimization; Scheduling and management of electric bus fleet; ET fleet feasibility\\
\bottomrule
\label{tab:assumption}
\end{tabular}
\end{table}

The first category concerns drivers' competing behavior in selecting charging station over the network. This assumption applies to the general EV modeling problem (both private and shared mobility) and is commonly found in  charging infrastructure planning problems, where user equilibrium is often assumed on the assignment rule for  charging route choice~\cite{he2014network, chen2016optimal, liu2017locating}. Specifically, the equilibrium flow assignment is established over the network to model EV users' charging behavior in response to the  planned charging facilities \cite{he2015deploying, chen2017deployment, chen2018cost}. Other studies also extend the notion of equilibrium to account for waiting time at charging locations in response to the capacity constraint\cite{chen2016optimal,chen2020optimal}. Under the similar setting, a more recent study~\cite{sun2020integrated} also models users' equilibrium responses to statisc and dynamic charging stations. In these studies, users are assumed to be selfish and perfectly rational,  seeking to maximize their individual utilities. Components in utility functions include travel time, energy consumption~\cite{zheng2017traffic}, waiting time for charging~\cite{chen2020optimal} and electricity cost~\cite{he2015deploying}. While user equilibrium presumes perfect information about these factors in a network, such a precondition is neither justified nor realistic in real-world cases. As a such, the resulting charging facility plans under the proposed framework are unlikely to be optimal in real-world setting. This undoubtedly prohibits the applicability for charging facility planning in large urban E-shared mobility systems. 

Under a non-competing setting (e.g., rule-based), the second category assumption is associated with drivers' preference on the location of charging stations. In these studies, spatial proximity is usually the major influencing factor and a typical assumption is that drivers always visit  nearest charging stations~\cite{ke2019modelling,song2019demand, tu2016optimizing}. In this regard, ET scheduling and fleet dispatching strategies highly depend on spatial distribution of travel demand and charging infrastructure.  Other studies also incorporate additional factors such as charging pile types, scale and supporting facilities at the charging stations~\cite{motoaki2017consumer,wolbertus2019electric}. Similar assumption is also observed  when exploring EV feasibility problems~\cite{hu2018analyzing} and also see its application in the charging facility planning problems~\cite{jia2012optimal,tu2016optimizing}. There are other studies assuming that ET drivers tend to charge at their own stations (home location)~\cite{ahn2016analysing, dong2017rec}, which further diversifies the possible outcomes regarding the driver location preference and charging decision-making. 

Besides drivers' choice for charging station location, assumptions on  drivers' choice for time of charging are also made in related studies in charging coordination strategy. It assumes that drivers' preferred charging time depends on the coordination among charging demand, charging price and future income. In particular, the overnight charging is assumed as the preferred charging time for lower charging and service fee~\cite{richardson2011optimal,masoum2016hybrid, heymann2017mapping}, which is also adopted in the scheduling and management of electric bus fleet~\cite{lajunen2018lifecycle,houbbadi2019optimal}. As for ET systems, the hypothesis that overnight charging will grant ETs sufficient range for one day of operation was often made when discussing ET feasibility~\cite{olsson2017lessons,moniot2019feasibility}. Since overnight charging enables ET drivers to replenish their vehicles off the shift, the daytime charging and fast charging option for ET systems are barely discussed. Nevertheless, the underlying assumption that all ETs will have access to overnight charging options is itself highly questionable and requires further justification. 


To summarize, the above three categories of assumptions are expected to play a decisive role in the E-shared mobility modeling process, and their validity will largely determine the effectiveness of decision-making in a real-world system. 
Given the limited practical evidence in the large-scale ET application, these assumptions have not been well justified and sufficiently debated. Therefore, there is a dire need of real-world evidence about E-shared mobility systems' charging behavior, which will establish the foundation for further planning and operation of  E-mobility systems.



\subsection{Studies on charging behavior with trip data}


We also note that there were several preliminary work on justifying the charging dynamics of E-shared mobility systems. With limited data sources, some studies attempted to explore the charging behavior of ET fleet through agent-based simulation~\cite{deyang2016simulation, sebastian2021simulation, yu2019spatial}. For instance, Bischoff et al.~\cite{bischoff2014agent} explored the performance of ET fleet and its feasibility through agent-based simulation. Yu et al.~\cite{yu2019spatial} analyzed the spatial-temporal charging pattern electric taxi based on trajectory data of conventional internal combustion taxis through using a simulation framework. Nevertheless, charging decisions in these studies were modeled based on the trajectory data of gasoline taxis, which is unlikely to represent he practical performance of E-mobility systems in the real world.
  
With the adoption of E-shared mobility in some big cities, few more recent studies attempted to investigate the charging behavior of ET fleet using the GPS trajectory data~\cite{tian2014understanding, rao2018modeling, fraile2018using}. However, we report that the contributions of these studies are restrictive due to limited ET fleet size and the coverage of service areas, which can only provide limited insights of charging and operating behavior of the E-shared mobility system. Specifically, Fraile-Ardanuy et al.~\cite{fraile2018using} evaluated the spatiotemporal energy usage pattern by summarizing the dynamics of 466 ETs in San Francisco. Tian et al.~\cite{tian2014understanding} explored the operating characteristics of 600 ETs in Shenzhen, China.  \added{Wang et al.~\cite{wang2020understanding} investigated the evolution of charging dynamics using ET trajectory data from 2013 to 2018, where the ET grew from 2.7\% to 65.2\% of the fleet, during which time the number of charging stations increased from fewer than 30 to 117. This study focused on the changes of mobility and charging patterns with increasing adoption of EVs. It provided important insights for staged planning, but of limited values for guiding the fully electrification of urban-scale fleet and long-term facility planning. Other researchers explored the driver behavior and charging characteristics of ETs or EVs based on realistic operational data including the State of Charge (SoC)~\cite{zou2016large,zhang2019usage,zhao2021assessment}. However, those papers concentrated more on the evaluation of energy consumption and efficiency, which is beyond our research scope. While the charging needs of ET systems highly relies on public charging infrastructures, a comprehensive understanding about the charging dynamics of fully electrified ET systems is critical for charging facility planning and ET operation management. Large-scale data set is therefore necessary to indicate the overall spatiotemporal distribution of charging activities. With the inevitable trend of fully electrification of urban shared mobility systems, a comprehensive understanding of fully electrified ET system is yet to be established.} 

So far, we have identified major gaps in the existing studies with respect to the charging behavior of E-shared mobility systems. Due to various restrictions, most studies failed to account for full complexity and subtlety of real charging behavior under a large-scale ET fleet. As such, efforts are made in this study to close existing gaps by exploring the spatiotemporal charging behavior under a large fleet of fully-electrified taxis from the system and individual levels. 

\section{Data}
\subsection{Electric taxi trajectory data}
The trajectory data used in this study cover the entire taxi fleet in Shenzhen, China, with a total of 21,371 vehicles. The ETs can be easily distinguished from the gasoline taxis based on the plate ID, and there are 20,130 ETs in total. This implies that Shenzhen has the largest ET fleet in the world, covering nearly 95\% of the entire fleet as in September 2019, which helps to gain insights into the charging dynamics of a fully electrified large-scale shared mobility service. There is only one model adopted by Shenzhen's ET fleet, BYD E6~\cite{byde62021}, with a battery capacity of 57 kWh and a mileage range of up to 300 km~\cite{wang2019sharedcharging}. The GPS trajectory data was obtained  through the collaboration with Shenzhen Transportation Planning Center. The data include the ET plate ID, longitude-latitude coordinates, the timestamps, and ET status (occupied or not).
An example of sampled ET trajectory data is shown in Table~\ref{table: data_example}. \added{It should be noted that although one month data might be relatively short to indicate long-term evolution of charging patterns, it is sufficient to provide valid observations in terms of daily patterns on system-level charging dynamics, the usage of charging infrastructure, and the differences in charging behavior at the individual level, which are concentration of the present work.} 

\begin{table}[ht!]
\small
\caption{Example data samples}
\begin{tabular}{llp{2cm}p{2cm}p{2cm}p{2cm}p{2cm}}
\toprule
\multirow{2}{*}{Taxi   GPS}       & plateID   & timestamp         & longitude & latitude                   & speed                      & occupied     \\ \cline{2-7} 
                                  & BDXXXX    & 9/18/2019   15:01:53 & 113.9745  & 22.55643                   & 42                         & 1            \\ \toprule
\multirow{2}{*}{Charging Station} & stationID & acessibility& longitude         & latitude  & fast-charging piles & slow-charging piles \\ \cline{2-7} 
                                  & 3892      & public& 114.342759        & 22.67359  & 15                         &5       \\
                                  \toprule
\multirow{3}{*}{Taxi schedule} & plateID & driver ID & shift  & primary shift days & secondary shift days & Total operation days\\ \cline{2-7} 
                                  & BDXXXX      & XX491 & Primary      & 30  & -    &30       \\
                                & BDXXXX      & XX191 & Secondary      & -  & 30    &30       \\\bottomrule
\end{tabular}
\label{table: data_example}
\end{table}


\subsection{Charging station information}

In addition to the trajectory data, we also collect the information of all charging stations in Shenzhen~\cite{shenzhen_charging_station2021}. Detailed information of these charging stations include station name, address, the number of slow-/fast-charging piles, charging service operator, accessibility (public or dedicated), open hours, electricity price, service fee, and parking fee. A sample of the collected charging station information is presented in Table~\ref{table: data_example}. We next match the addresses with the longitude-latitude coordinates via the python package `Geocoder'~\cite{geocoder2021}. To better describe different types of zone, we categorize the City of Shenzhen into three classes: \textit{city center, suburb, and periphery}, based on the traffic analysis zone. This classification criteria is analogous to Nie's work~\cite{nie2017can}, where a detailed layout of Shenzhen is included therein.

\begin{figure}[H]
    \includegraphics[width = 0.9\linewidth]{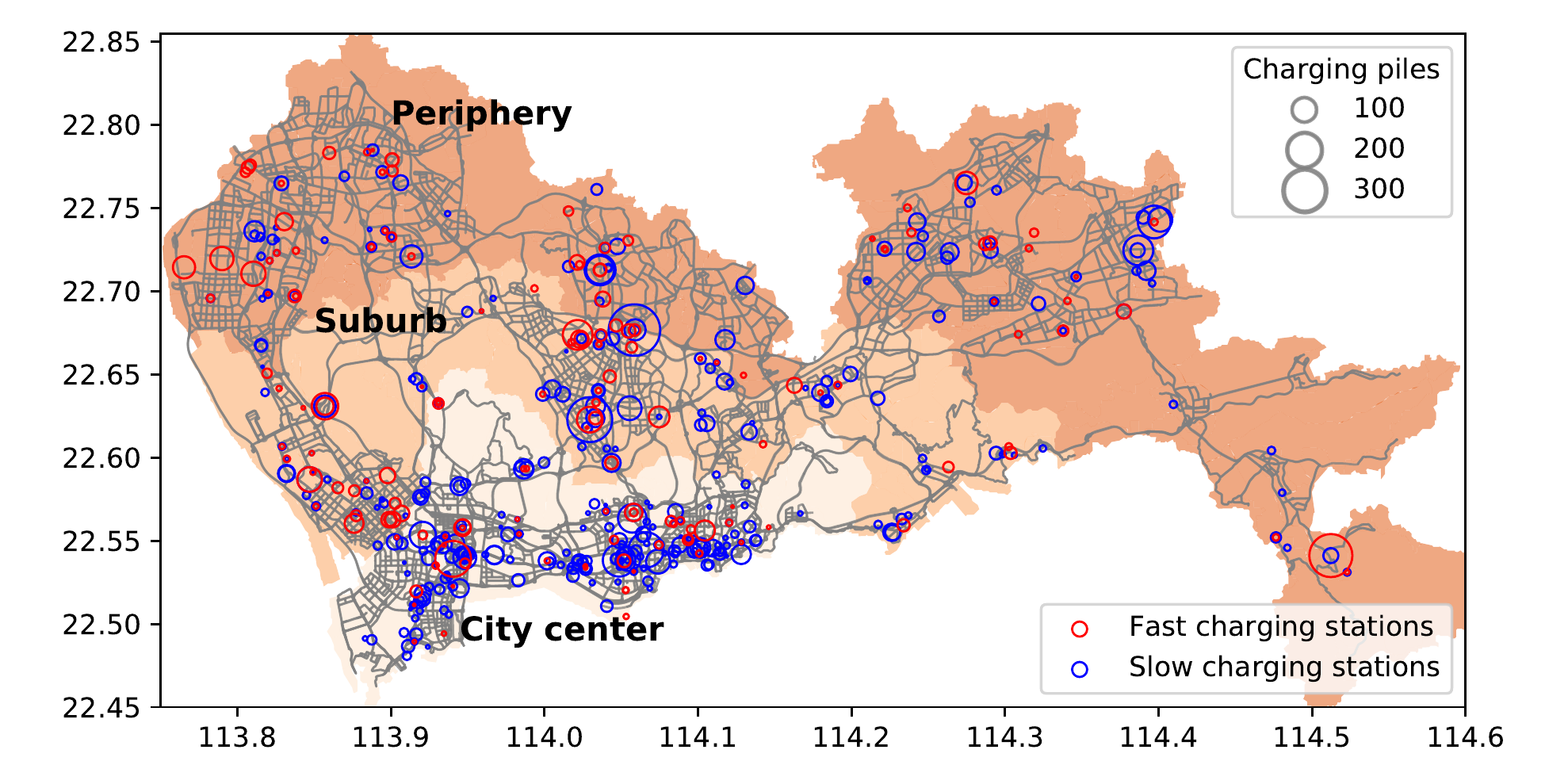}
    \caption{Distribution of slow and fast charging stations with the radius of the circle representing the number of corresponding charging piles.}
    \label{fig:cs_distribution}
\end{figure}

After removing private charging stations and charging stations without charging piles, we end up with 425 charging stations in total. Following the standards adopted in the China EV market, we use the terms "slow charging" and "fast charging" to represent Level 2 charging and DC Fast charging, respectively~\cite{zheng2020systematic}. The slow charging indicates the pile with power less than 10 kW (industrial commons are either 3.5 kW or 7 kW). And the fast charging pile can at least charge 80\% of a 50kWh ET within one hour under the charging rate of up to 60 kW~\cite{hu2018china}. Among the 425 charging stations, \deleted{there are} 160 of them are installed with fast charging piles. The spatial distribution of these charging stations, along with the number of charging piles, are shown in Figure~\ref{fig:cs_distribution}. It can be observed that the charging stations are more densely located \added{in} the city centers of Shenzhen (including Nanshan, Futian, and Luohu district), most of which are small-sized stations (e.g., fewer than 100 piles) with primarily slow charging piles. On the other hand, there are a few large-scale charging stations with more than 300 charging piles and they mainly located on the edge of the core business districts or in more distant peripheral areas.

\subsection{Taxi driver shift schedule}
We also obtain the taxi drivers' shift schedule in September 2019, which include information about the ET company, plate ID, the drivers on primary and secondary shift, and the total operation days per driver. A preprocessing of the data suggested that there were 19,943 active ETs in this month, accouting for over 99\% of all ETs in Shenzhen (20,130 in total). We next divide the ETs into two sets based on their operation mode: single-shift and double-shift ETs, which accounts for 33.2\% (6,622 total ETs) and 66.6\% (13,321 total ETs) respectively. As for the ET drivers, there are 35,549 individual drivers: 4,630 of whom served in single-shift mode and the other 30,919 drivers served in double-shift mode. As a consequence, double-shift is the dominant mode for daily ET operations. In the following sections, we will further explore the charging behavior of urban ETs considering the difference between the two shift modes. 

\section{Method}
\subsection{Charging event identification}
\begin{figure}[H]
    \centering
    \includegraphics[width=\linewidth]{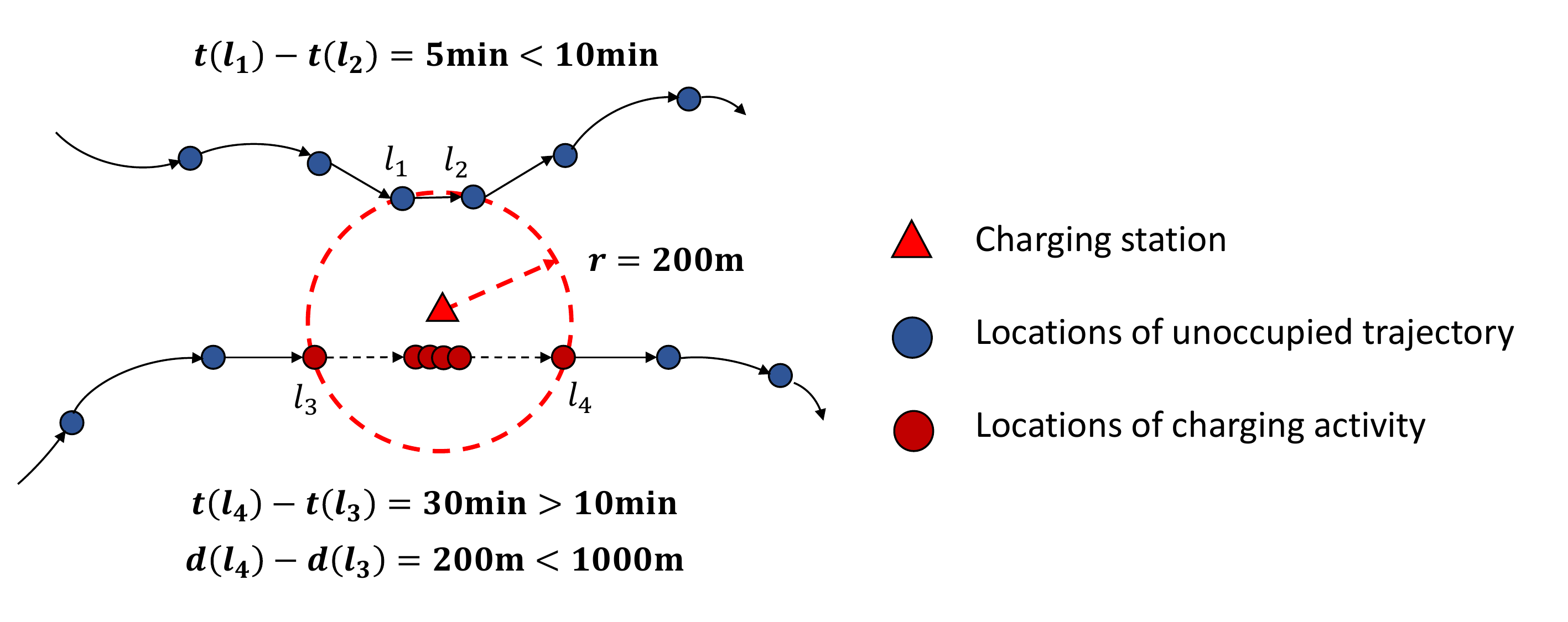}
    \caption{Illustration of charging event identification}
    \label{fig:illustration_identify}
\end{figure}

To explore the charging behavior of ETs, we first identify charging events from the trajectory data based on the proximity to the charging stations and the dwelling duration. We consider that a valid charging event would require a minimum stop duration. This can be translated into first finding stopping events: a sequence of trajectory points with a minimal change in the GPS location and the corresponding time duration exceeds a user specified threshold. Besides, the taxi should be unoccupied following the loading status of each ET in the data. Moreover, it should be noted that the identified stopping events should be adjacent to a charging station. Based on these conditions, charging events are then identified using the following rules and an example is shown in Figure~\ref{fig:illustration_identify}. 

First, we define a stopping event to be valid if (1) the total spatial displacement does not exceed \added{1,000 meters} for at least \added{10 minutes ($\Delta d \le 1000$ and $\Delta t \ge 10$)}, and (2) the nearest charging station is located within \added{200 meters ($\Delta d_{CS} \le 200$)}. In addition, the ET drivers may turn off their GPS device for multiple times during the operation horizon (e.g., taking a break, having lunch, etc.).
As such, we also consider the potential charging event while the GPS device is turned off. Let $n_{off}$ and $n_{on}$ be the last location that the GPS device is off and the first  location that the GPS device is on. Specifically, if the GPS is off for \added{more than 10 minutes}, we only require that (1) the displacement is under \added{1,000 meters} and (2) either $n_{off}$ or $n_{on}$ is within \added{200 meters} to the nearest charging station ($\Delta d^{cs}_{n_{off}}\le 200$ or $\Delta d^{cs}_{n_{on}}\le 200$).

After processing, each piece of identified charging event is represented as a data tuple, consisting of (1) the start and end time of charging, (2) location coordinate of charging, (3) the ID of the nearest charging station, (4) the travel time and distance to the charging station since the last passenger being dropped off, and (5) the travel time and distance from the charging station to the next passenger being picked up. 

\textbf{Ground truth}
To further examine the feasibility of charging event identification method, we take two different approaches to verify the validity of the identified charging events. 

First, we conducted a field investigation to reach out to taxi drivers in Shenzhen (by street \& App hailing) and obtained their help in distributing an online survey to the drivers’ community. The survey was designed using a tool called Wenjuanxing (Chinese version of Qualtrics) and mainly inquired charging related behavior. To increase the return rate, we intended to send out a short survey with only 6 questions, and the details of the survey can be found in the Appendix. Survey questions are about ET drivers' charging frequency, minimum and maximum stay time, and the preferred charging behavior. After one week period, 172 valid responses were collected from 55 single-shift (32\%) drivers and 117 double-shift drivers (68\%). The population is biased towards single-shift drivers as 14\% of all drivers are single-shift per the official record. We then visualize below the charging frequencies of both types of drivers, and the results suggested that single-shift drivers on average charge 1.6 times a day while double-shift drivers will charge more than 1.7 times a day. The weighted average daily charging frequency across all ETs is 1.7 times.

\begin{figure}[H]
    \centering
    \includegraphics[width=0.5\linewidth]{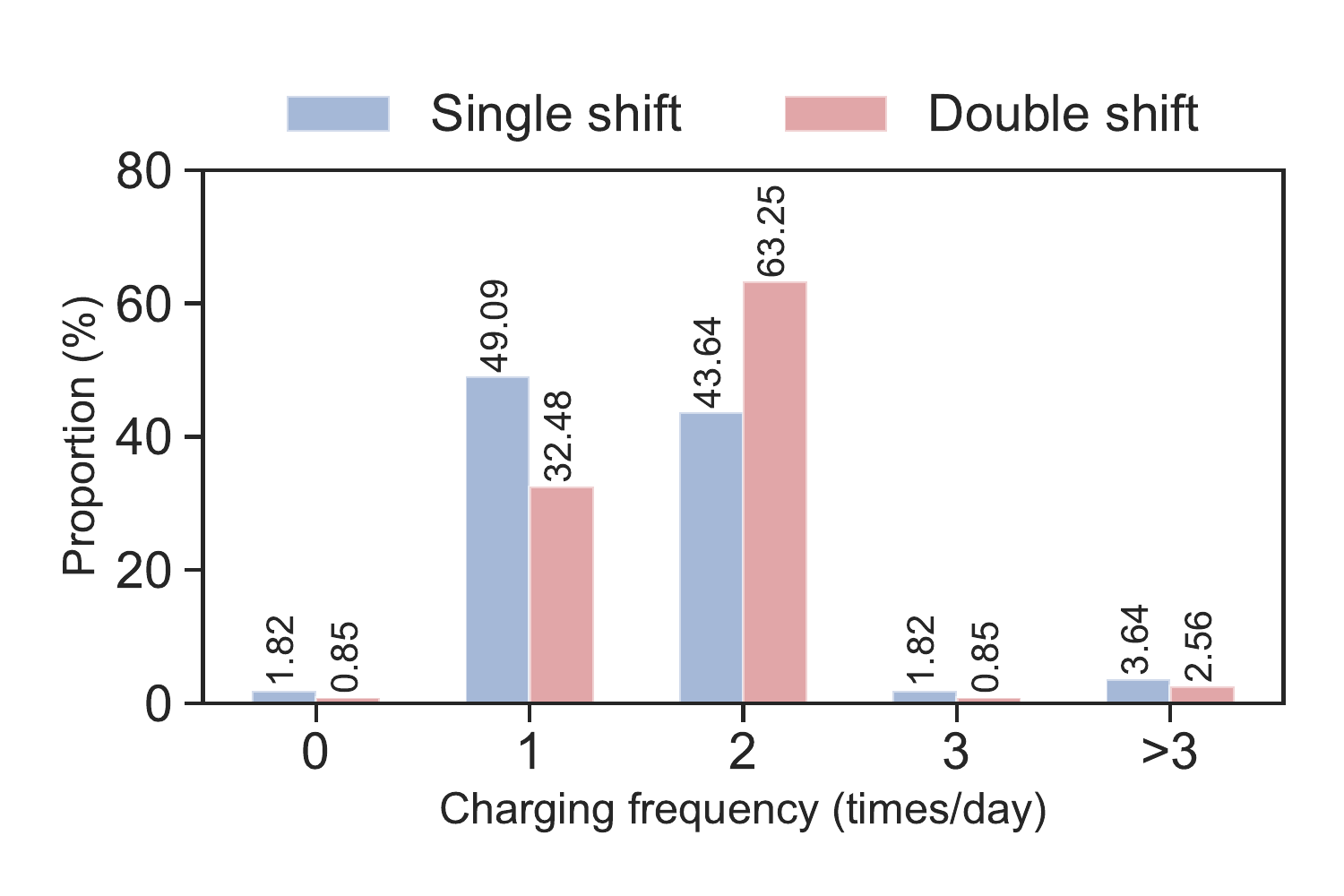}
    \caption{Survey statistics about ET drivers' charging frequency}
    \label{fig:charge_fre_mode}
\end{figure}

\added{
Besides the charging frequency, the minimum and maximum stay time at a charging station are especially helpful in determining the hyperparameters for charging events identification. We show below the distribution of minimum and maximum stay duration for both types of drivers. It is observed most of the drivers (95\%) stay a minimum of 15 minutes at the charging stations and the charging duration is usually less than 3 hours. Single-shift drivers, on the other hand, may stay longer to fully charge their battery (11\%).   Our survey results are different from previous studies where a 30-minute time threshold was used for charging events identification from ET GPS trajectory data~\cite{tian2014understanding,wang2020understanding}.The reason for such a difference is likely due to the advances in fast charging facilities in recent years. }

\begin{figure}[H]
    \centering
    \subfloat[Minimum stay time at charging stations by shift]
    {\includegraphics[width=0.5\linewidth]
    {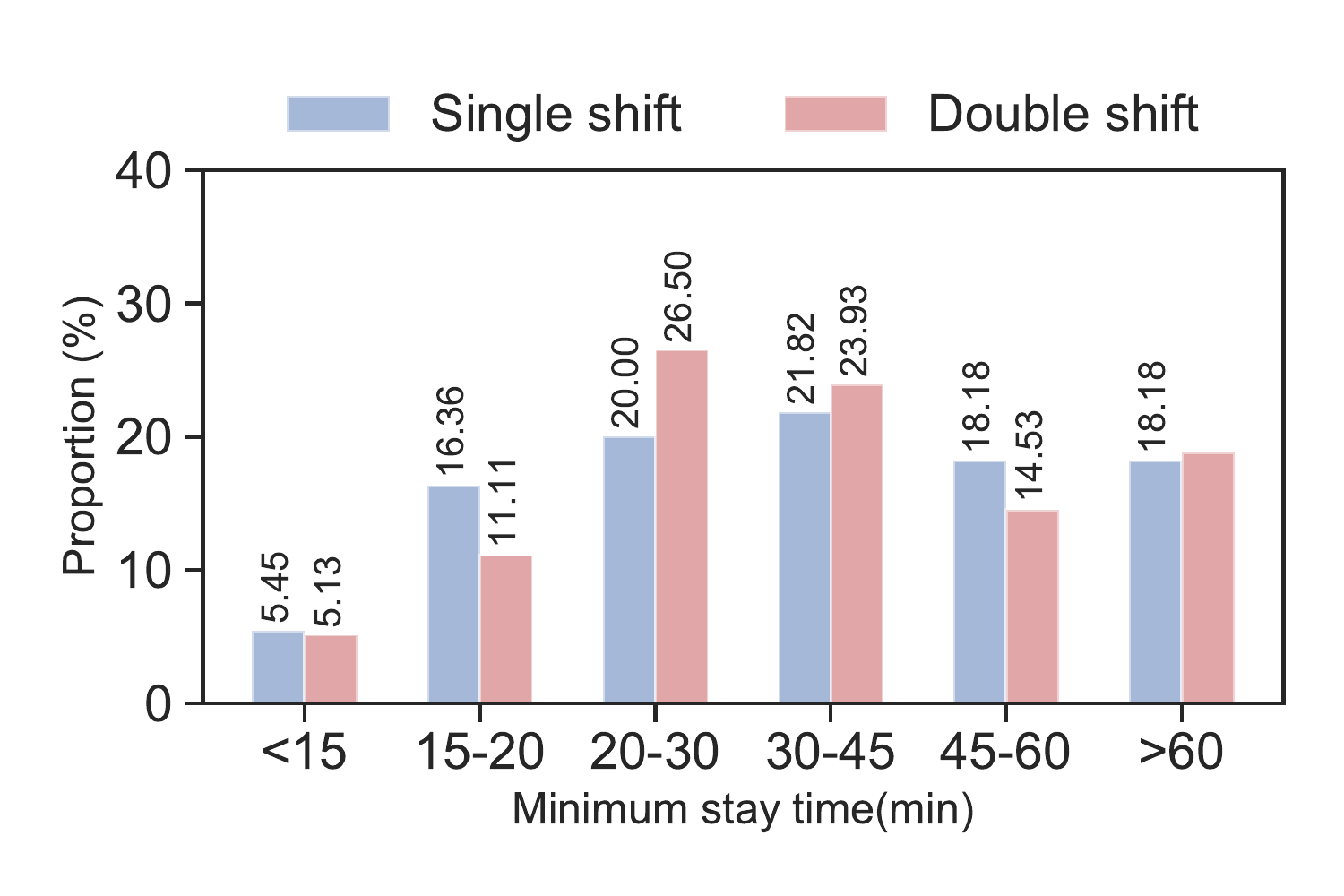}
    \label{fig:charge_min_mode}}
    \subfloat[Maximum stay time at charging stations by shift]
    {\includegraphics[width=0.5\linewidth]
    {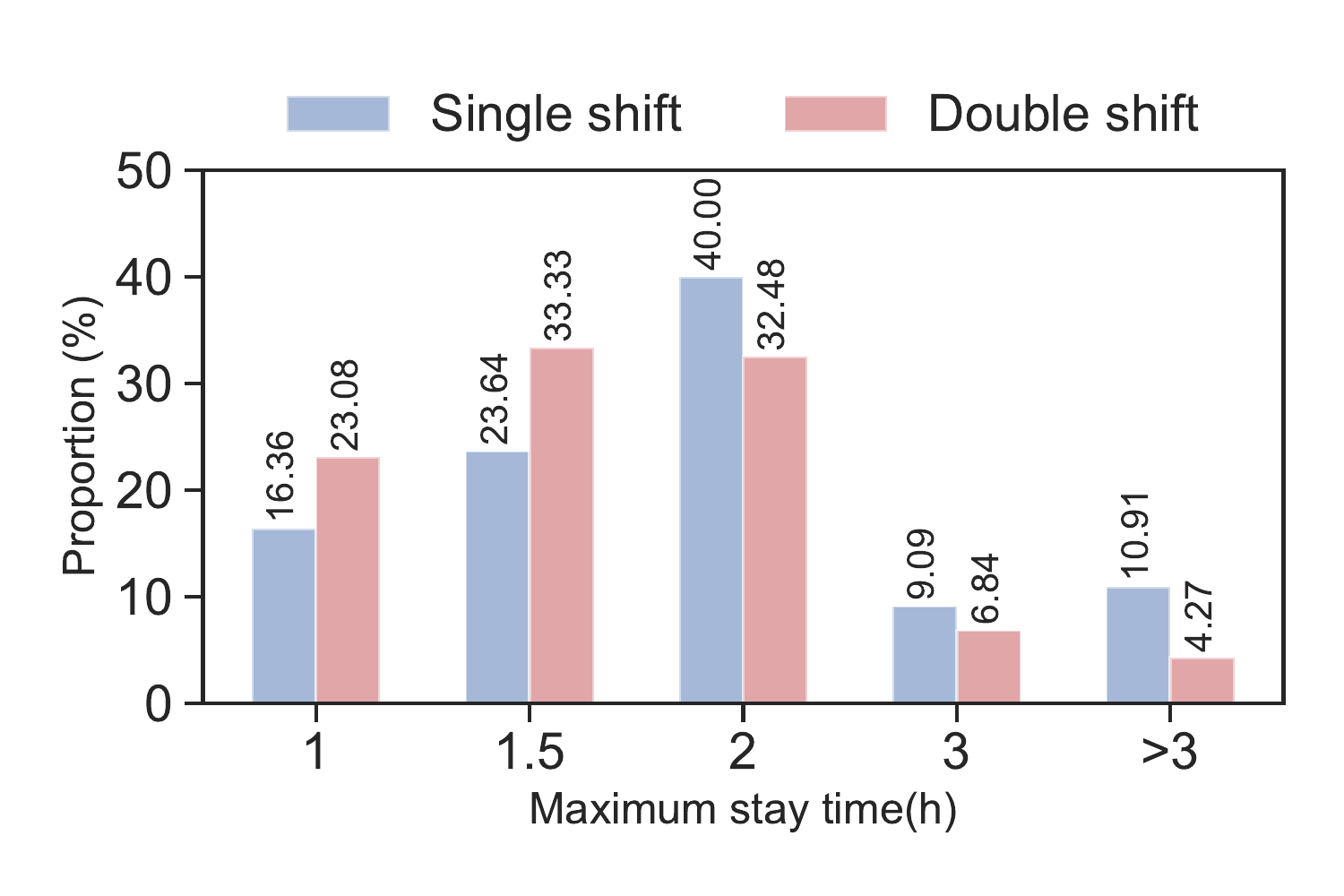}
    \label{fig:charge_max_mode}}
    \caption{Survey statistics about time spent at charging stations}
    \label{fig:charge_times}
\end{figure}

\begin{figure}[H]
    \centering
    \includegraphics[width=0.9\linewidth]{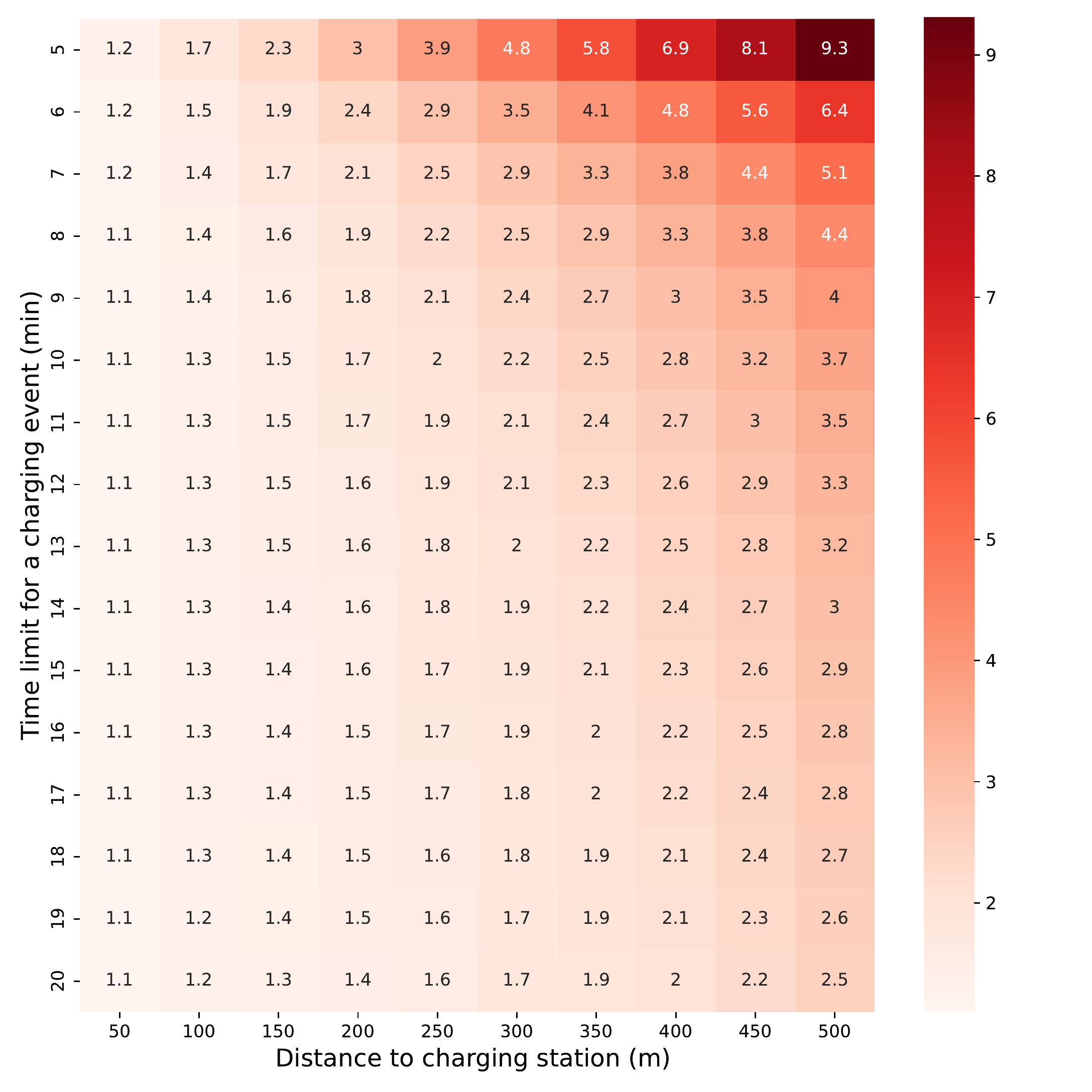}
    \caption{Average charging times per ET under different justification rules}
    \label{fig:ave_charging_times}
\end{figure}

\added{
Second, based on the survey results, we further conducted a cross-validation experiment to verify how the number of charging events is affected by the distance threshold and time threshold. The results are presented in Figure~\ref{fig:ave_charging_times}. In detail, 16 temporal thresholds and ten distance limits are considered, resulting in a grid search of 160 scenarios. The temporal thresholds to identify a sufficient charging duration are from 5min to 20min, incremented by 1min. And the spatial thresholds for the distance limit to the nearest charging station are from 50 to 500, incremented by 50. The results indicate that a combination of from 200 meters and 10 minutes to 300 meters and 15 minutes may all be appropriate for charging events identification, with the average number of identified charging events ranging from 1.7
to 1.9 per ET. By combining the findings from both the survey and the cross-validation experiment, we decide to choose a minimum time threshold of 10 minutes which is capable of capturing almost all charging events. And the distance threshold is then set to 200 so that the average number of charging events is 1.7. This combination best agrees with our survey results with the weighted average number of charging frequency being 1.7, and is also aligned with the findings in previous studies~\cite{wang2020understanding,zou2016large}.
}

\subsection{Spatiotemporal charging behavior}
With the identified charging events, we next perform spatial-temporal analyses to understand the consistency of charging behavior at the system level. 

First, to investigate the temporal dynamics of daily charging behavior, we use the data from the \added{first four weeks} of September and group the city-level charging events into one-hour time intervals. The autocorrelation analysis is then conducted using the constructed time series data to examine the distribution of autocorrelation values over different lag values. The autocorrelation value lies in the range of -1 to 1, with 1 indicating a strong positive correlation and -1 for strong negative correlation at the corresponding time lag $k$. Denoting the number of charging events at time $t$ as $Y_{t}$,  the number of charging events at time $t+k$ as $Y_{t+k}$ and the average number of charging events as $\bar{Y}$, the autocorrelation at lag $k$ can be computed as: 
 
 \begin{equation}
    \gamma_{k} =\frac{\sum_{t=0}^{T-k}  (Y_t-\bar{Y})(Y_{t+k}-\bar{Y})}{\sum_{t=0}^T{(Y_t-\bar{Y})^2}}
\end{equation}

To understand the spatial dynamics of everyday charging events, we further group the charging events at each individual charging stations for every one-hour time interval as the representation of spatial distribution for charging events. This creates a vector $C_t$ of length $1\times425$ for every time interval $t$. We then calculate the Pearson correlation between $C_t$ and $C_{t+k}$ to evaluate if there are significant differences between the spatial distributions of two time intervals. In particular, the Pearson correlation coefficient can be measured as:

\begin{equation}
    \gamma(t,t+k) = \frac{\sum_i (C_t^i-\bar{C}_t)(C_{t+k}^i-\bar{C}_{t+k})}{\sqrt{\sum_i(C_t^i-\bar{C}_t)^2\sum_i(C_{t+k}^i-\bar{C}_{t+k})^2}}
\end{equation}
where $C_t^i$ denotes the $i$th element of $C_t$ and $\bar{C}_t$ represents the average of all elements in $C_t$. One benefit of Pearson correlation is that the results are scale invariate. In this regard, even though the total number of charging events may not be strictly the same between two time intervals, a Pearson correlation value close to 1 will still imply a highly consistent distribution of charging activities at different charging stations.

\subsection{Modeling individual charging behavior}

We model individual charging behavior using a two-dimensional Latent Dirichlet Allocation (LDA) model, which was first introduced by Blei et al.~\cite{blei2003latent} and extended by Sun et al.~\cite{sun2021routine} to capture the spatiotemporal charging patterns of individual ET drivers. The LDA method is an unsupervised method that helps to identify latent topics in the large corpus of documents, which can be analogous to obtain classes of charging patterns from the sequence of charging events. In our study, we consider each charging event as a spatiotemporal point. Let $\mathcal{S}$ be the set of charging stations for the spatial dimension and $\mathcal{T}$ represent the set of hourly intervals of one day. In this regard, the charging events of an ET can be expressed as follows:

\begin{equation}
    E_{v} = \{\left(e_{vi}^{s},e_{vi}^{t}\right): i \in \{1, ..., N_{v}\}, s \in \mathcal{S}, t \in \mathcal{T}\}
\end{equation}
where $E_{v}$ is the collection of charging events for all ETs $v\in\mathcal{V}$. $N_{v}$ is the total number of charging events for ET $v$ in our data. The tuple $(e_{vi}^{s},e_{vi}^{t})$ demonstrates the charging events on spatial and temporal dimensions. Note that the spatial points $e_{vi}^{s}$ are labeled by the charging station IDs regardless of their exact coordinates, since we focus more on finding the representative spatial patterns than the neighborhood clusters based on physical distance. To better understand our framework, we compare the notions used in this study with the typical elements in the LDA model: topics, documents, and words (or set of vocabulary). In detail, we consider charging patterns ($z\in\mathcal{Z}$) as \textit{topics}, ETs $v\in\mathcal{V}$ as \textit{documents}, and charging events $(e_{vi}^{s},e_{vi}^{t})$ as \textit{words}. The major purpose of utilizing the two-dimension model is to simultaneously capture the spatial and temporal activity patterns on the individual level while considering their dependencies~\cite{sun2021routine}. Compared with user-level analyses based on the high spatial-temporal dimension by enumerating $\mathcal{S}\times\mathcal{T}$~\cite{hasan2014urban}, the two-dimension model exhibits the merits under the significantly varying size of the spatial and temporal dimensions ($|\mathcal{S}|= 425$,$|\mathcal{T}|=24$ in our case).

\begin{figure}
    \centering
    \includegraphics[width=0.85\linewidth]{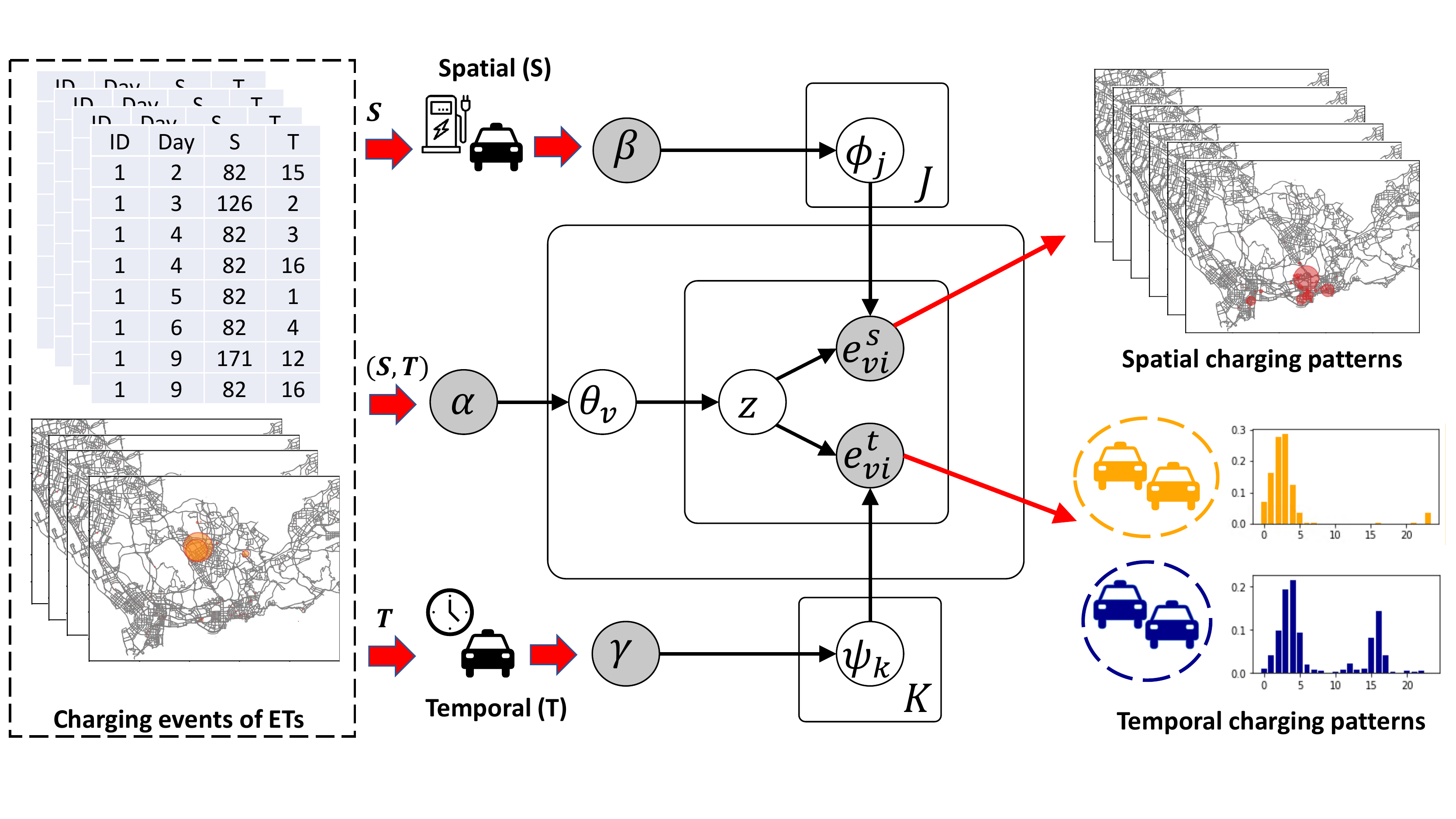}
    \caption{Graphical model of the two-dimensional LDA adapted from~\cite{sun2021routine}}
    \label{fig:lda_model}
\end{figure}

We present the graphical model of our LDA in Figure~\ref{fig:lda_model}. The parameters in the grey circles can be either pre-determined (i.e., $\alpha,\beta,\gamma$) or observed in data (i.e., $e_{vi}^{s},e_{vi}^{t}$). Let $J$ and $K$ denote the number of patterns on spatial and temporal dimensions. In this case, $\phi_{j}$ is the $j$-th spatial pattern-event distribution and $\psi_{k}$ is the $k$-th temporal pattern-event distribution. Further, the charging pattern distribution $\theta_{v}$ can be expressed a two-dimensional simplex, denoted by $J\times K$, with respect to the probability distributions on spatial and temporal dimensions. We specifically express such two-dimensional simplex as $\theta_{vjk}$ such that $0\le\theta_{vjk}\le1$ and $\sum_{j=1}^{J}\sum_{k=1}^{K}\theta_{vjk}=1$. Formally, the generative process in the LDA model~\cite{steyvers2007probabilistic} can be expressed as follows.

\begin{enumerate}
    \item[\textbf{Step 1}]: For each ET $v\in\mathcal{V}$, select a pattern distribution $\theta_{v} \sim Dirichlet_{|\mathcal{S}|\times|\mathcal{T}|}\left(\alpha\right)$
    \item[\textbf{Step 2}]: For each spatial pattern select spatial distribution $\phi_{j} \sim Dirichlet_{|\mathcal{S}|}\left(\beta\right)$
    \item[\textbf{Step 3}]: For each temporal pattern select time distribution $\psi_{k} \sim Dirichlet_{|\mathcal{T}|}\left(\gamma\right)$
    \item[\textbf{Step 4}]: For each ET $v\in\mathcal{V}$, for each charging event $\left(e_{vi}^{s},e_{vi}^{t}\right)$:
        \begin{enumerate}
            \item Select one pattern $z$ from the charging events following $z = \left(z^{s},z^{t}\right) \sim M(\theta_{v})$
            \item Select one charging event on spatial dimension $e^{s}$ given the pattern $z^{s}$ following $e^{s} \sim M(\phi_{z^{s}})$
            \item Select one charging event on temporal dimension $e^{t}$ given the pattern $z^{t}$ following $e^{t} \sim M(\psi_{z^{t}})$
        \end{enumerate}
\end{enumerate}

In the generative process, $\alpha$, $\beta$, and $\gamma$ are the Dirichlet priors. And $M$ is the Multinomial distribution. It is worthy to note that the Dirichlet distributions are selected as the prior since they are conjugate to the Multinomial distribution. By integrating $\theta_{v}$, $\phi_{j}$, and $\psi_{k}$, we can compute the joint distribution $P(e_{vi}^{s},e_{vi}^{t})$, which can be expressed as:

\begin{equation}
    P(e_{vi}^{s},e_{vi}^{t},z^{s},z^{t}) = P(e_{vi}^{s}|z^{s})P(e_{vi}^{t}|z^{t})P(z^{s},z^{t})
\end{equation}
where the terms $P(e_{vi}^{s}|z^{s})$, $P(e_{vi}^{t}|z^{t})$, and $P(z^{s},z^{t})$ can be obtained following similar forms as in Griffiths and Steyvers~\cite{griffiths2004finding}.
However, there is no closed-form solution to directly obtain the posterior distribution for each ET $v$, following $P(z^{s},z^{t}|e_{vi}^{s},e_{vi}^{t}) = \frac{P(e_{vi}^{s},e_{vi}^{t},z^{s},z^{t})}{\sum_{s=j}^{J}\sum_{t=k}^{K}P(e_{vi}^{s},e_{vi}^{t},z^{s},z^{t})}$. Instead, the parameters can be estimated under Markov Chain Monte Carlo procedure using Gibbs sampling approach~\cite{geman1984stochastic}. Readers are referred to Griffiths and Steyvers~\cite{griffiths2004finding} for a detailed description on the parameter estimation using Gibbs sampling, and Sun et al.~\cite{sun2021routine} for the modeling on spatiotemporal activity patterns. 

\subsubsection{Model selection}
In the LDA model, the number of spatial and temporal charging patterns $J$ and $K$ are pre-determined. We adopt the perplexity to find the optimal combination of $J$ and $K$, where the lower perplexity score indicates better performance on the model generalization. The procedure of calculating the perplexity score are as follows. Given the trained model, we select a set of ETs as the validation set. And the perplexity of ET $v$ can be expressed as below:

\begin{equation}
    perp(\mathbf{e}_{v}^{s},\mathbf{e}_{v}^{t}|\mathbf{e}_{train}) = \exp{\left(-\frac{\log P(\mathbf{e}_{v}^{s},\mathbf{e}_{v}^{t}|\mathbf{e}_{train})}{N_v}\right)}
\end{equation}
where $P(\mathbf{e}_{v}^{s},\mathbf{e}_{v}^{t}|\mathbf{e}_{train})$ can be approximated by Monte Carlo method~\cite{sun2021routine}, which can be expressed as an average over the samples in validation set $\mathcal{U}$:

\begin{equation}
    P(\mathbf{e}_{v}^{s},\mathbf{e}_{v}^{t}|\mathbf{e}_{train}) = \frac{1}{|\mathcal{U}|} \sum_{u\in\mathcal{U}}\prod_{i=1}^{N_v}\left(\sum_{j=1}^{J}\sum_{k=1}^{K} \theta_{vjk}\phi_{e_{vi}^{s}~j}^{u}\psi_{e_{vi}^{t}~k}^{u} \right)
\end{equation}

Given the perplexity score, we will conduct a grid search procedure to determine the best combination of $J$ and $K$ with the lowest perplexity score.

\subsection{Individual driver regularity} 



After introducing the topic model in identifying charging patterns, we then investigate the potential regularity of individual charging activities. Specifically, the temporal regularity for ET $v\in\mathcal{V}$ for all weeks $w\in\mathcal{W}$ can be quantified by the cosine similarity. The cosine similarity is scale-invariant and bounded between 0 and 1, which makes it an ideal choice for comparing regularity among ET drivers. We first denote by $\cos(w,w^{\prime}|v)$ the cosine similarity for ET $v$ between week $w$ and $w^{\prime}$, which follows the form:

\begin{equation}
    \cos\left(w,w^{\prime}|v\right) = \frac{\mathbf{p}(z^{s}, z^{t}|e_{vi}^{s,w},e_{vi}^{t,w})\cdot \mathbf{p}(z^{s}, z^{t}|e_{vi}^{s,w^{\prime}},e_{vi}^{t,w^{\prime}})}{||\mathbf{p}(z^{s}, z^{t}|e_{vi}^{s,w},e_{vi}^{t,w})|| ~ ||\mathbf{p}(z^{s}, z^{t}|e_{vi}^{s,w^{\prime}},e_{vi}^{t,w^{\prime}})||}
    \label{eq:cosine_similarity}
\end{equation}
where $\theta_{vjk}^{w}$ can be obtained following the similar procedures as described, by treating $\mathbf{e}_{v}^{w,s},\mathbf{e}_{v}^{w,t}$ as a dedicated set of charging events for ET $v$ in week $w$. We next compute the expectation of the ET $v$'s cosine similarity between all weeks $w\in\mathcal{W}$, which can be expressed as follows:

\begin{equation}
    \overline{\cos\left(v\right)} = \frac{1}{|\mathcal{W}\times\mathcal{W}|} \sum_{w\in\mathcal{W}}\sum_{w^{\prime}\in\mathcal{W}} \left(\cos(w,w^{\prime}|v)\right)
    \label{eq:expectation_document_in_topic}
\end{equation}








\section{Results}
\subsection{System-level charging dynamics}

\begin{figure}[H]
    \subfloat[Daily charging events and number of unique ETs]{\includegraphics[width=0.5\linewidth]{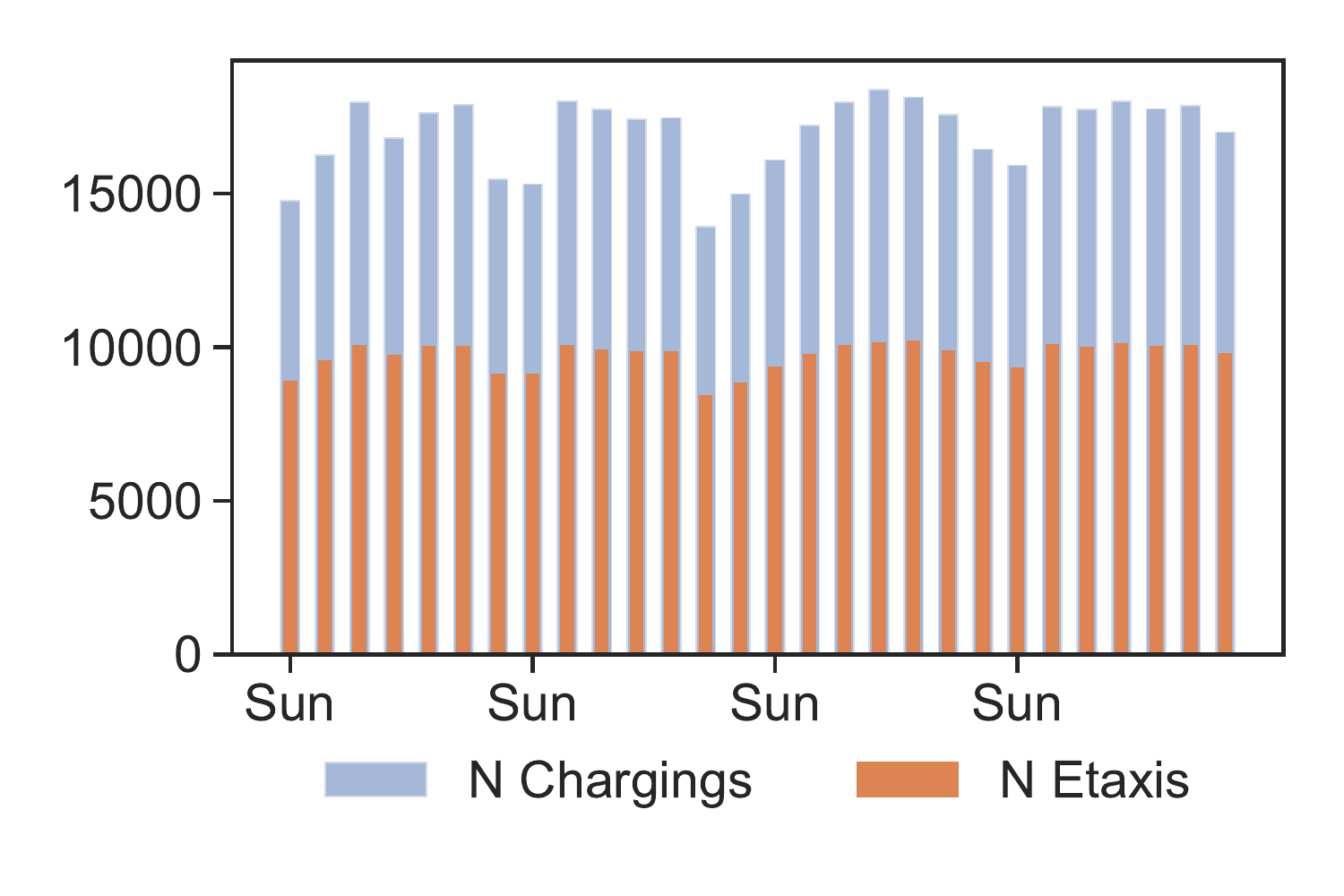} \label{fig:N_events_drivers}}
    \subfloat[ET captured by number of days]{\includegraphics[width=0.5\linewidth]{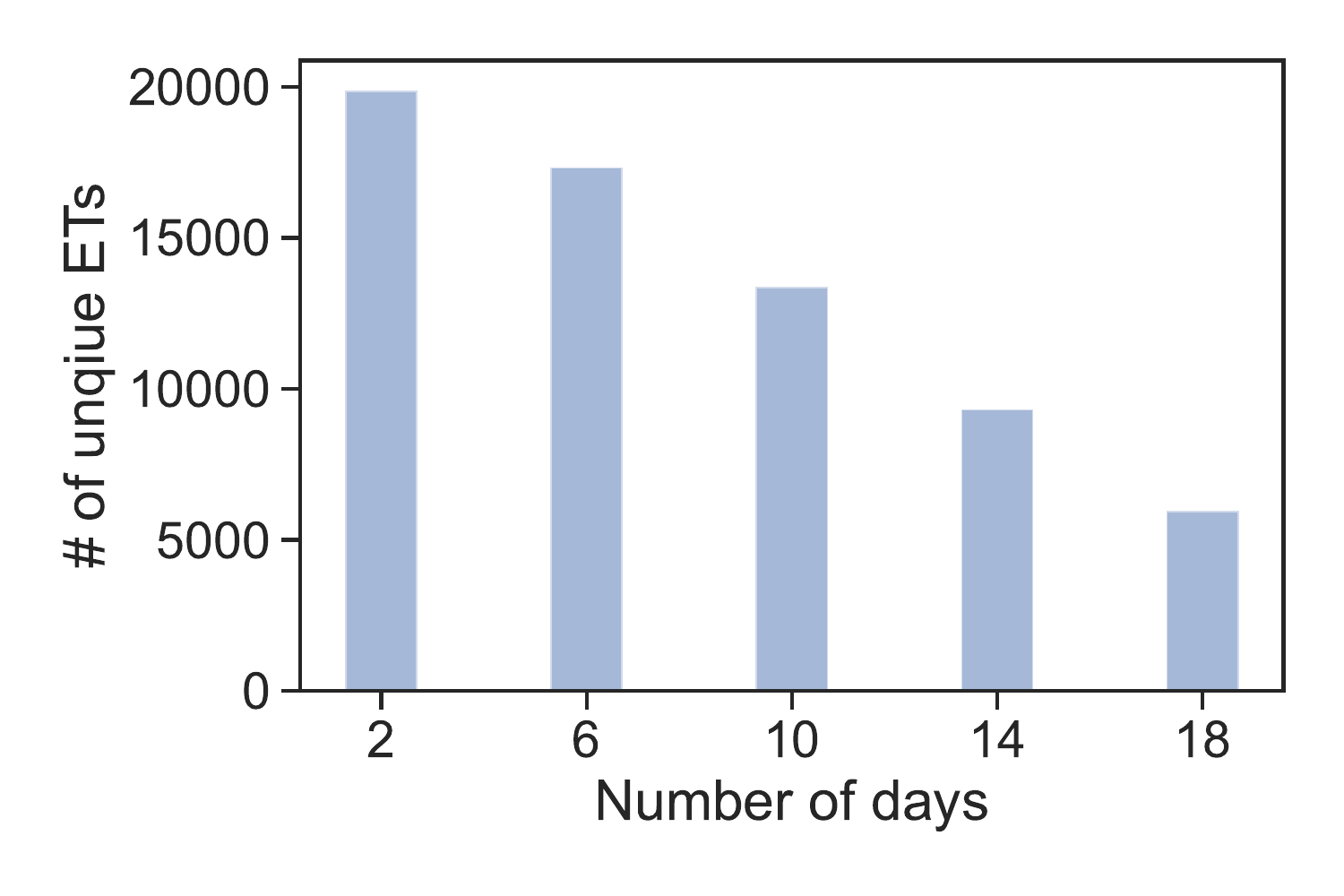}\label{fig:vehicle_covered}}
    \caption{Number of identified charging events and number of ETs covered in a selected week}
    \label{fig:weekly_pattern}
\end{figure}

With the identified charging events, we first report the distribution of the number of charging events, the number of unique ETs \added{(the ETs with the same plate ID)} and accumulated number of unique ETs over the first four weeks (9/1/2019 - 9/28/2019) as shown in Figure~\ref{fig:weekly_pattern}. Note the sample size on 9/13 (Friday) is comparably smaller due to the Mid-Autumn Festival with a one-day break. As seen in Figure~\ref{fig:N_events_drivers}, the daily-average number of charging events is \added{17,015} and the average number of unique ETs covered by the charging events is \added{9,725}, or nearly 48.6\% of all ETs. As a result, the identified daily number of charging events cover fewer number of ETs than the actual fleet size and the reasons are multi-fold. First, not all ETs may operate on a daily basis and \deleted{the not all ETs may} require a daily charging, especially those on a single shift. Second, some of the ETs may turn on and off their GPS devices outside the vicinity of a charging station and are therefore not captured by our identification algorithm. Finally, GPS device errors could be another reason. Nevertheless, the identified charging events still capture a stable overall trends over the entire week for majority of the ETs, with Sunday having slightly fewer number of charging activities as compared to other days of the week. 



On the other hand, we also find that there are \added{19,880} ETs that appears at least two days over the four-week period (see Figure~\ref{fig:vehicle_covered}, accounting for over 97.5\% of total operated ETs in this month. The number of ETs that are captured for at least 6 and 10 days of the week are \added{17,336 and 13,370}, respectively. This implies that the data contain sufficient details so that longitudinal behavior at the individual level can be tracked for over \added{66.5}\% of all ETs in the city. Finally, even in the case when more than 18 days are required, the identified events can still cover \added{29.9\%} unique ETs, which is significantly greater than any reported cases in the literature. Based on these observations, we argue that the charging events identified are \added{reliable}, and the number of charging activities and unique ETs covered are sufficient to derive statistically representative conclusions for the entire ET population.

\begin{figure}[H]
    \includegraphics[width=\linewidth]{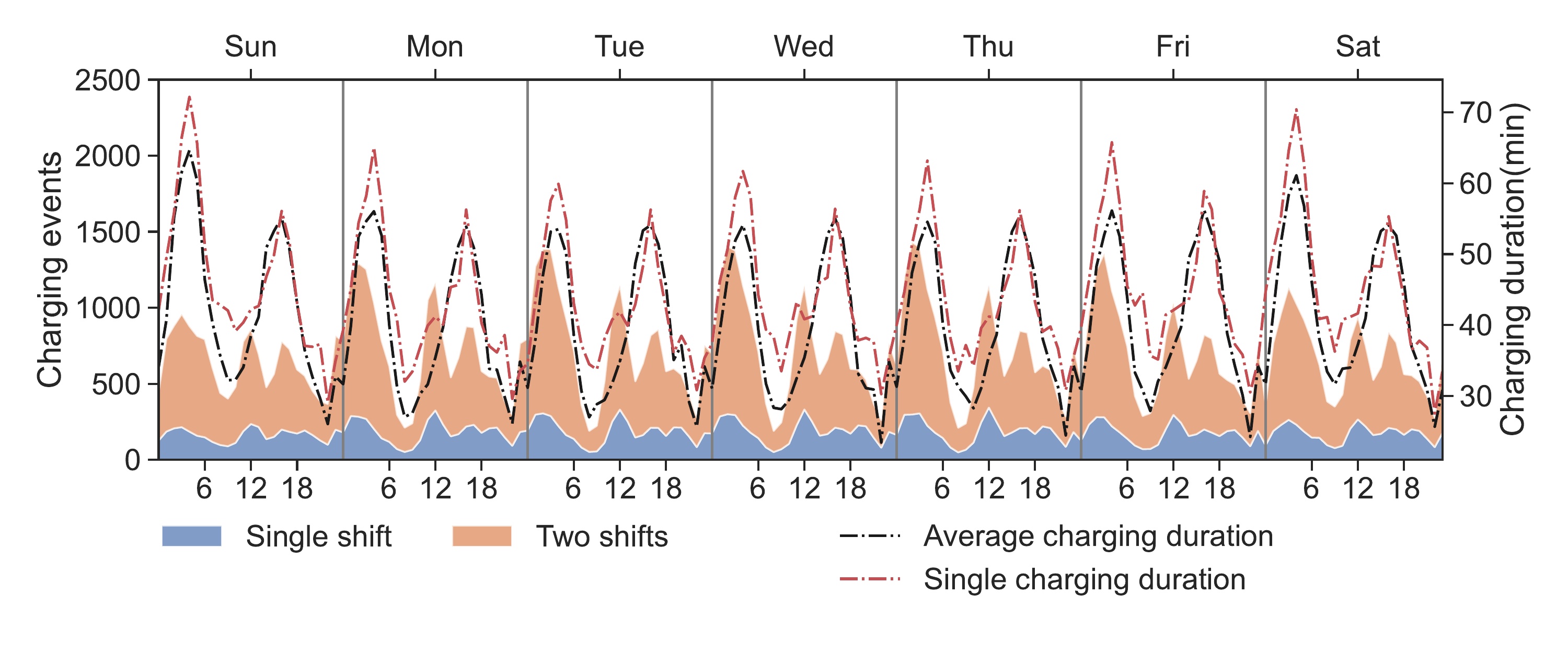}
    \caption{Temporal distribution of charging dynamics based on drivers number of shifts.}
    \label{fig:charging_shift}
\end{figure}

After verifying the quality of the identified charging events, we next \added{concentrate on the weekly-average observations and }investigate the within-day dynamics of ET charging and examine system-level differences between single-shift and double-shift ETs. Figure~\ref{fig:charging_shift} shows the temporal distribution of charging events for both single-shift and double-shift ETs on the weekly average. Besides, the average charging duration for charging events that took place in the corresponding hour is also visualized. One immediate observation is that the daily charging dynamics at the system level show repetitive patterns over the week, which will be further investigated in the following discussions. In addition, we found that the total number of charging events for double-shift ETs is significantly higher than that of single-shift ETs. While the total number of double-shift ETs doubles that of single-shift ETs, the number of charging events for double-shift vehicles is more than three times higher than that of single-shift ETs at most time of the day. This clearly articulates the difference in within-day charging dynamics between the two types of ETs and implies that their behavior should be treated differently when modeling the ET services. Despite the differences in charging frequency, both classes of ETs follow a similar pattern of three daily charging peaks, taking place at late night period (from 2 AM to 5 AM), noontime period (from 11 AM to 1 PM ), and late afternoon to the evening time period (from 4 PM to 7 PM), respectively. These three charging peaks correspond to four distinct types of charging behavior. The noontime charging is aligned with the lunch break so that ET drivers can save dedicated charging trips, and such type of charging behavior is observed for both single and double-shift ETs. Moreover, the charging duration is found to be significantly shorter (around 30 minutes shorter on average) than most of the other time periods. The only exception is around 8 PM, which may correspond to the time of the dinner break. The afternoon period is aligned with the afternoon shift schedule of double-shift drivers, and this type of behavior is more evident among double-shift drivers. Finally, the late-night charging activities are found to be different between single and double-shift ETs. In particular, the peak for single shift ETs occurs 1-2 hours earlier than double shift ETs while having a longer charging duration. For double shift ETs, this corresponds to the morning shift schedule, where the drivers again need to be responsible for the electricity consumption of their night shift before turning the ET to the next driver. As for single-shift ETs, the observations mainly capture the behavior of those drivers who do not have access to charging infrastructure at home, and they may need to visit the charging station before ending or starting their shift. 


\begin{figure}[H]
    \includegraphics[width=.8\linewidth]{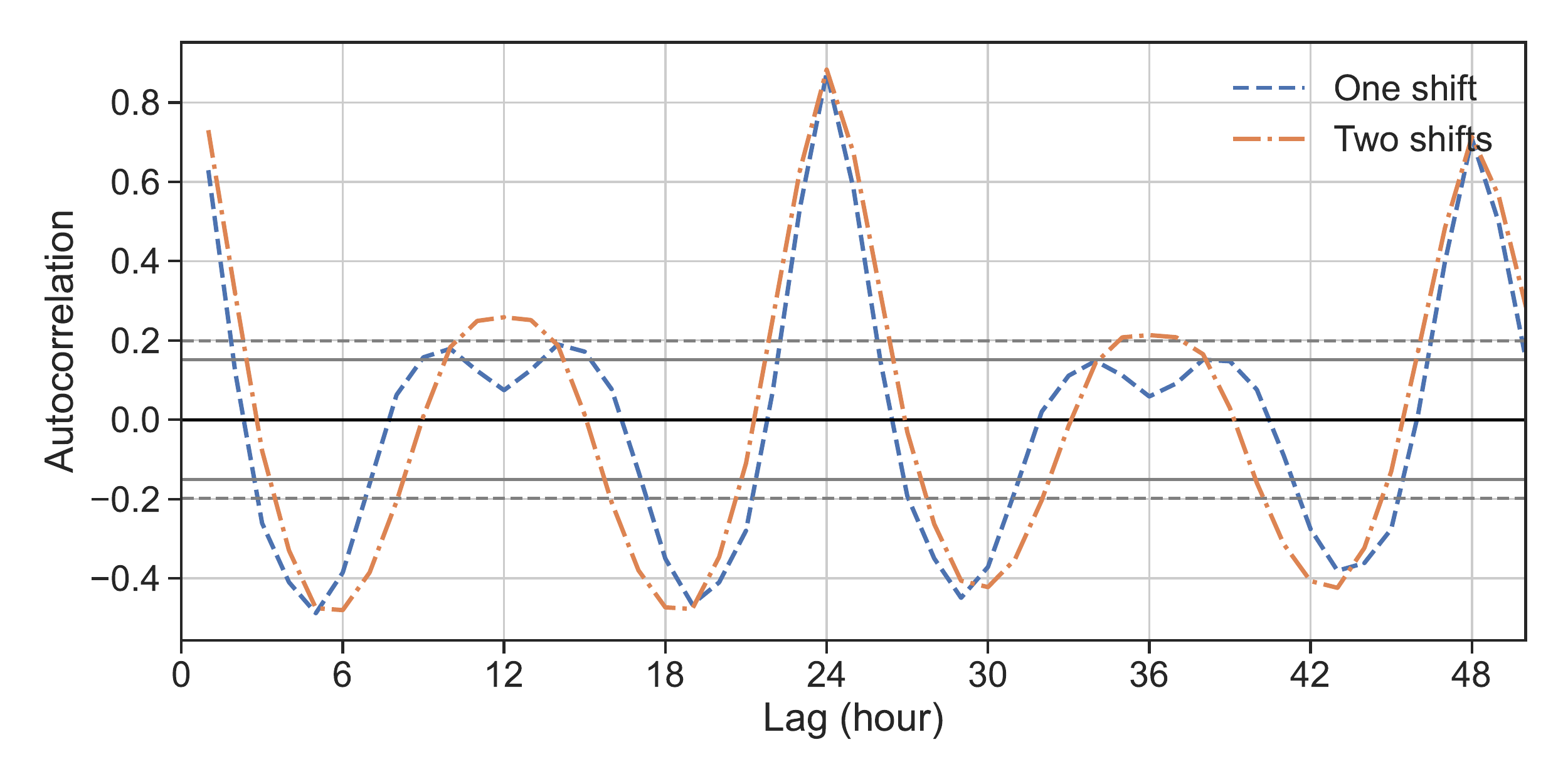}
    \caption{Temporal autocorrelation of charging events at system level}
    \label{fig:time_corr}
\end{figure}

While observing a trend of repeating dynamics in Figure~\ref{fig:charging_shift}, we next quantitatively evaluate the temporal consistency of dynamics. The autocorrelation is computed with the time series of charging counts at the hourly level. And the autocorrelation plot for both single-shift and double-shift ETs is shown in Figure~\ref{fig:time_corr}. The 90\% and 95\% confidence levels are also visualized as solid and dash lines, respectively. Based on the results, a significant temporal correlation that exceeds the lines of 95\% confidence interval can be observed at different time lags. Notable time lags of interest are located around lags 6, 18, 24, 30, 42, and 48.
Most importantly, the autocorrelation takes the highest value at the time lag of 24, followed by the time lag at 48. With each lag being a one-hour time interval, the results present statistically significant evidence that the system level charging dynamics are stable and highly consistent on a daily basis. And this observation holds for both single and double shift ETs despite their different within-day dynamics. On the other hand, strong temporal cross-correlations are also observed every 6 and 18 hours. These two values are reflective of the shift schedules for double shift ETs, where each driver is responsible for a shift of 12 hours, and the results again echo the findings in Figure~\ref{fig:charging_shift}.

\begin{figure}[H]
    \includegraphics[width=.8\linewidth]{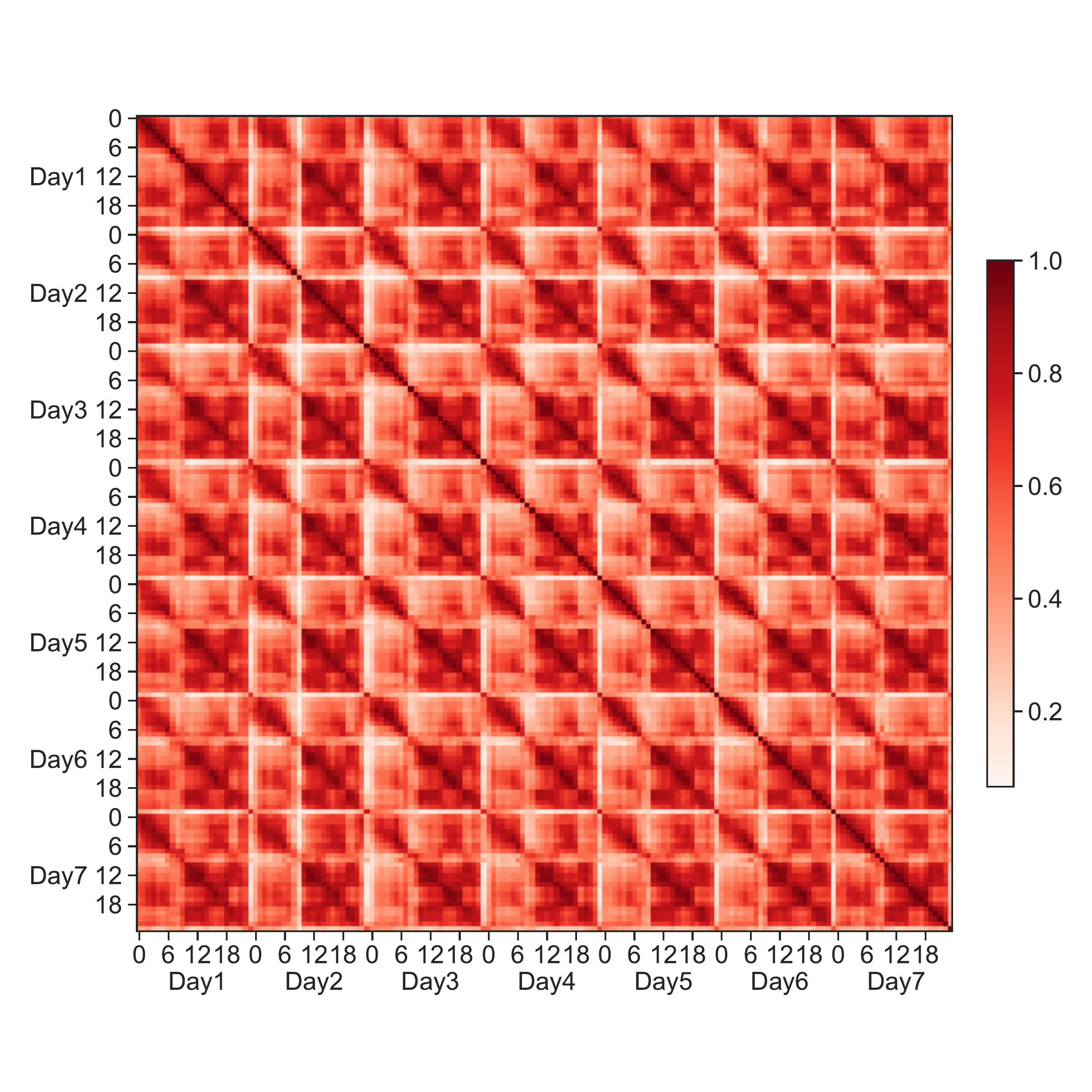}
    \caption{Pearson correlation of charging station usage at different hours of day}
    \label{fig:spatial_corr}
\end{figure}

Besides the temporal autocorrelation, a more striking finding on system-level charging dynamics can be found through the Pearson correlation between the spatial charging activity distributions of two time intervals. Figure~\ref{fig:spatial_corr} presents a $168\times168$ matrix summarizing the pairwise Pearson correlation coefficients over the entire week. While daily regularity was reported at the aggregate level, such a regularity still holds even considering the charging counts at individual charging stations. That is, a Pearson correlation coefficient close to 1 is still captured every 24 hours. And further details can also be understood by tracing local blocks of high Pearson correlation coefficients. Specifically, local blocks of high correlation can be found near 12 PM, between 0-6 AM, between 4-7 PM, as well as from 8 PM to 11 PM.
As a consequence, besides regularity on a daily basis, the charging dynamics can be categorized into at least four clusters based on the local affinity of charging activity distributions at individual charging stations, where the charging dynamics are self-similar within these time intervals. \added{Finally, we find two time windows, one between 9-11 AM and the other shortly before midnight, where the charging activity distributions are most dissimilar to all other time intervals (where the value of Pearson correlation is close to zero and is shown as the white space) over the day. Moreover, it is observed from Figure ~\ref{fig:charging_shift} that these two time windows also correspond to the lowest number of charging activities across the day, which further implies that the charging behavior is least structured and more ad-hoc during the off-peak time.} 

\subsection{Dynamics at charging stations}
With strong spatial and temporal regularity at the system level, we next discuss how charging events are distributed at individual charging stations.

\begin{figure}[H]
    \includegraphics[width=0.6\linewidth]{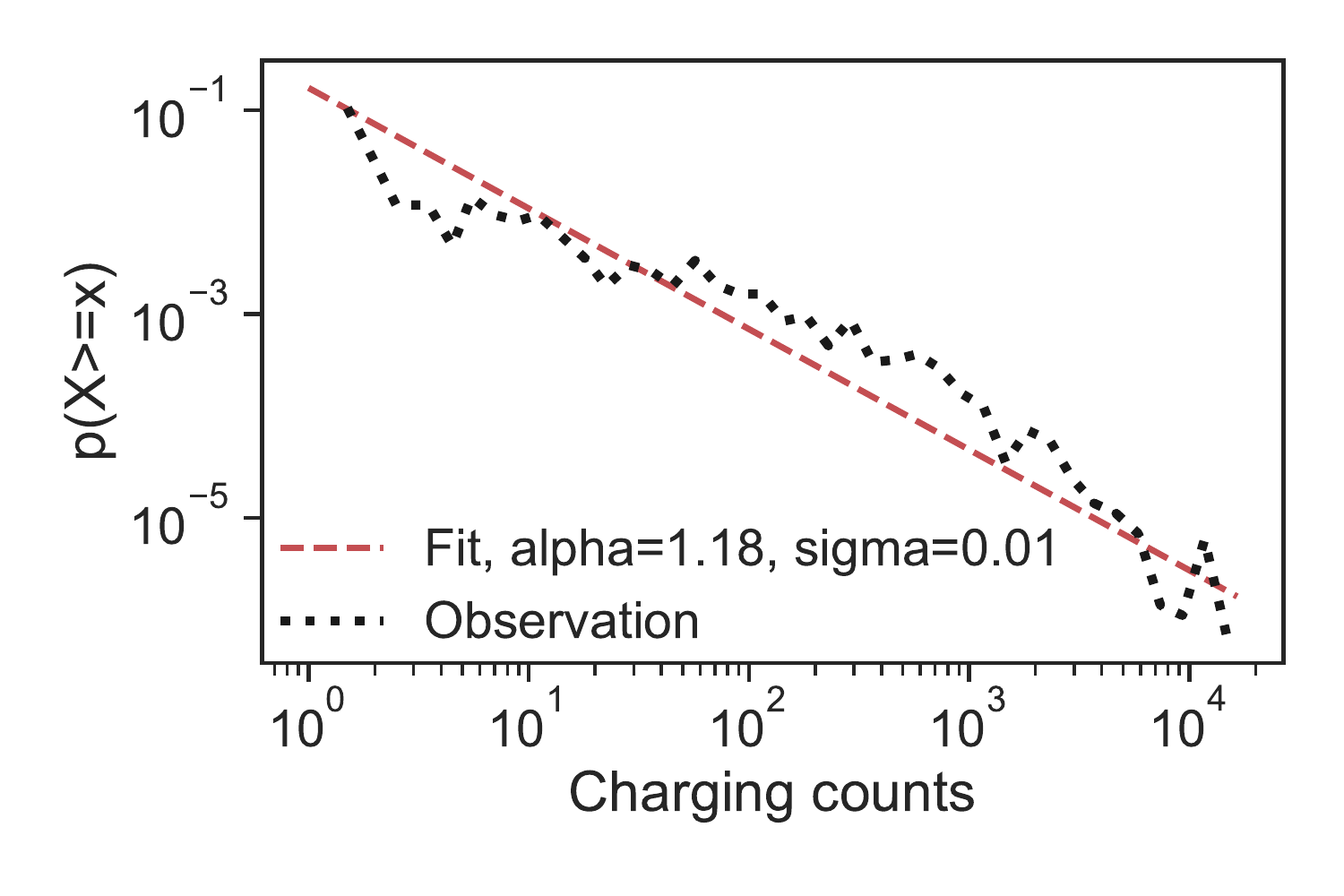}
    \caption{CCDF of charging visits at all charging stations in log-log scale.}
    \label{fig:power_law_charging}
\end{figure}

Figure~\ref{fig:power_law_charging} shows the daily-average charging frequency of all 425 stations. One important observation here is that the complementary cumulative distribution function (CCDF) of charging frequency across the stations can be well approximated by the power-law distribution $P(X\geq x)\propto x^{-\alpha}$, with $\alpha$ calibrated as \added{1.18}. As such, the station-level usage shows a strong trend of preferential attachment, indicating that a small fraction of the charging stations serves majority of the charging demand (\added{24 stations or less than 6\% of all stations} account for more than 50\% of total charging demand). In particular, based on the statistics at each station, we report that \added{45} charging stations are found to have never been visited, \added{136} stations are visited 1 to 10 times daily, and \added{149} stations are visited 10 to 100 times. Only \added{29} charging stations (12\%) are more frequently visited with more than 100 daily charging activities, where 3 of them are much more heavily preferred by the drivers with over \added{600} daily visits. These observations lead to significant implications in understanding the charging preference of ET drivers. On the one hand, the charging station choice is unlikely to be ad-hoc. Instead, the charging behavior is highly structured and is likely to be planned ahead for most of the drivers, where they prefer to visit stations that they are familiar with or stations that are frequently used by other drivers. On the other hand, the power-law distribution also alerts the resilience issue for satisfying the charging needs of large-scale ETs. The power failure at top visited charging stations may result in significant disturbances to the ET mobility services as they are searching alternate charging stations. This asserts the need for considering resilience and back-up plans when planning charging facilities.

\begin{figure}[H]
    \subfloat[Spatial distribution of charging station usage]{\includegraphics[width=\linewidth]{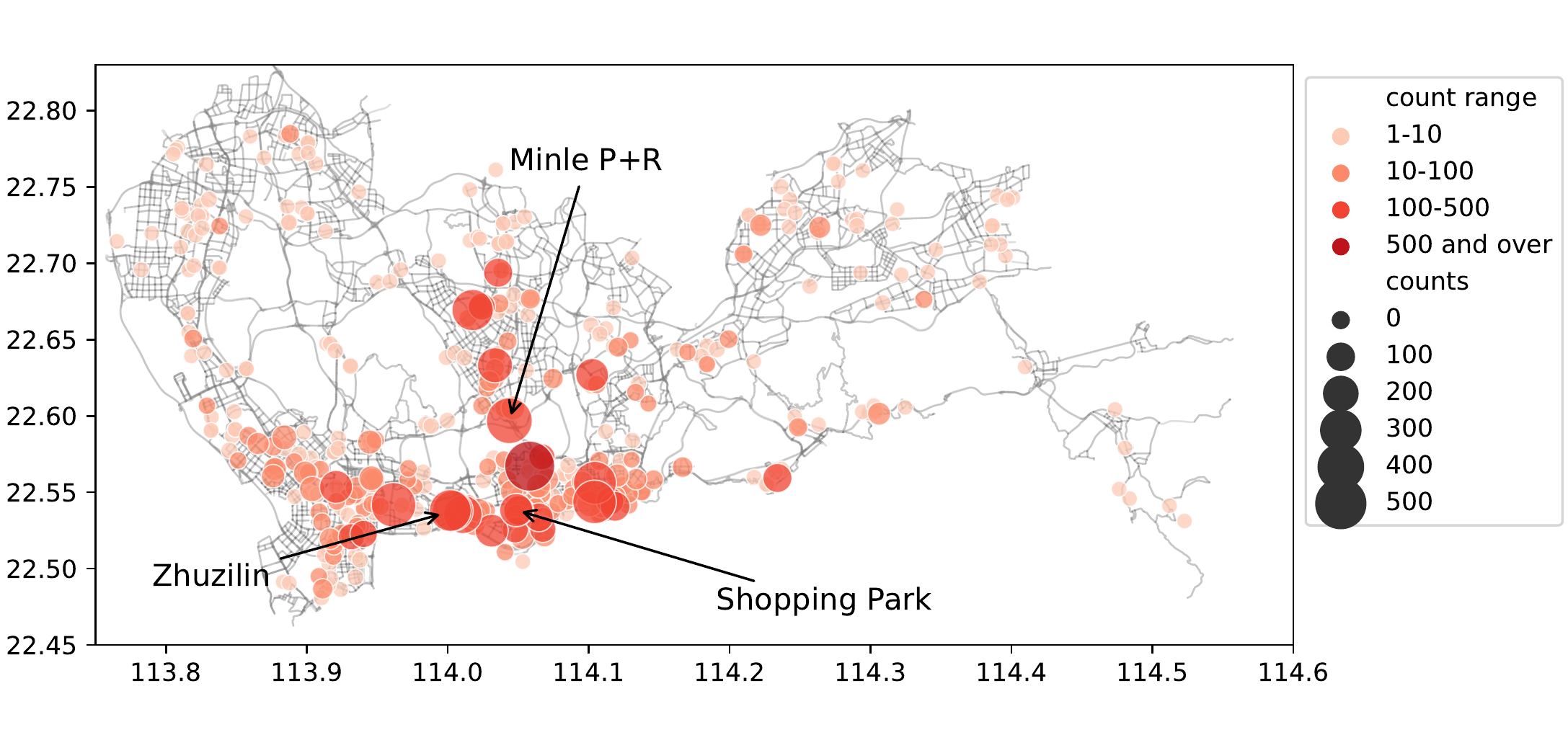}\label{fig:spatial_cs}}\\ 
    \subfloat[Minle P+R station]{\includegraphics[width=0.3\linewidth]{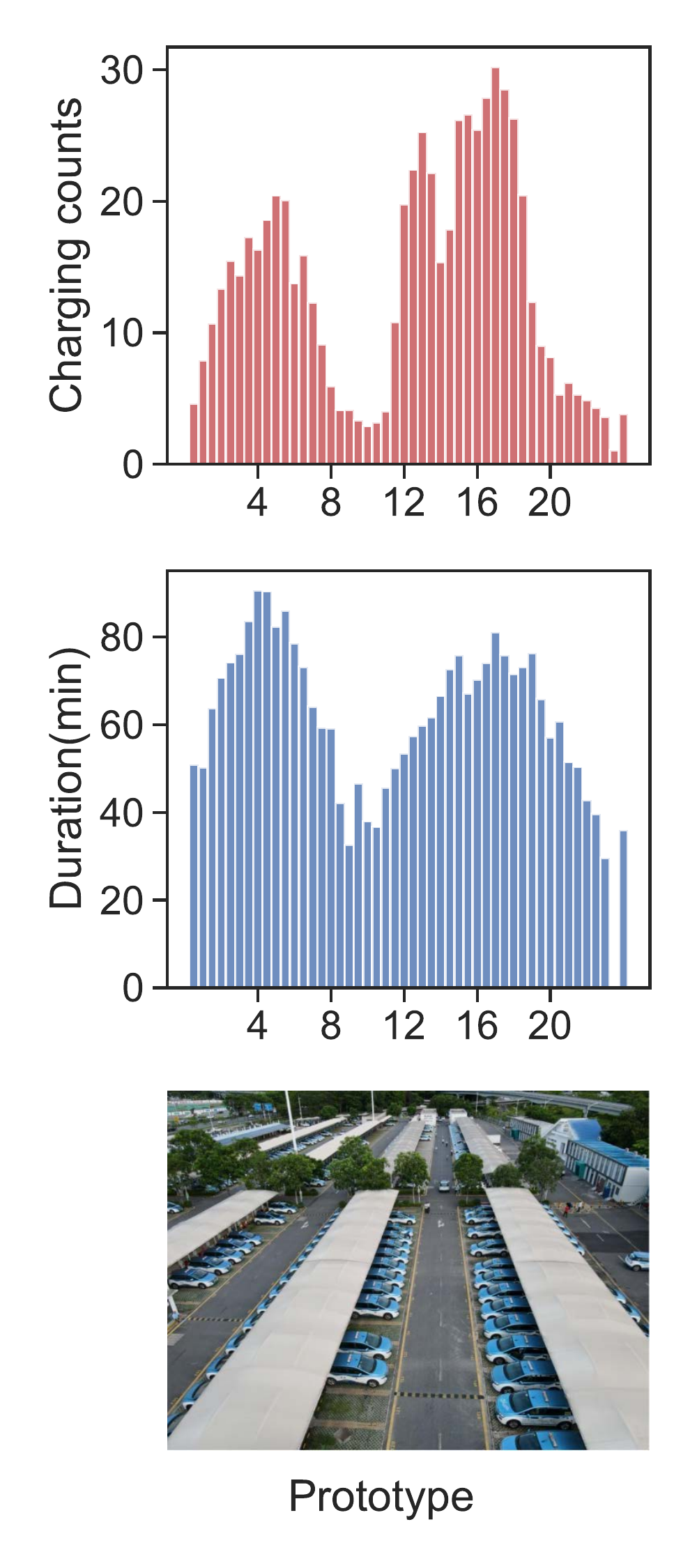}\label{fig:cs_minle}}
    \subfloat[Zhuzilin station]{\includegraphics[width=0.3\linewidth]{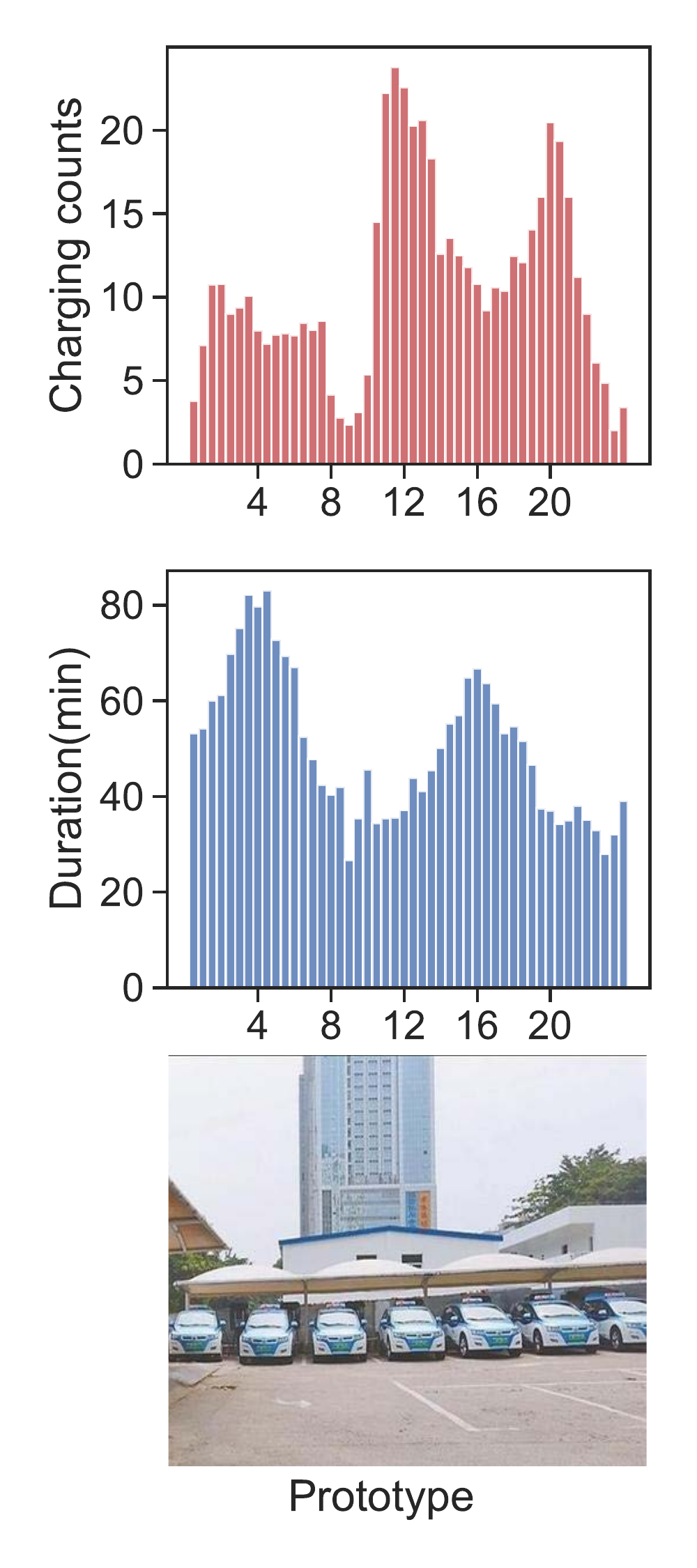}\label{fig:cs_zhuzilin}} 
    \label{fig:spatial_distribution}
    \subfloat[Shopping Park station]{\includegraphics[width=0.3\linewidth]{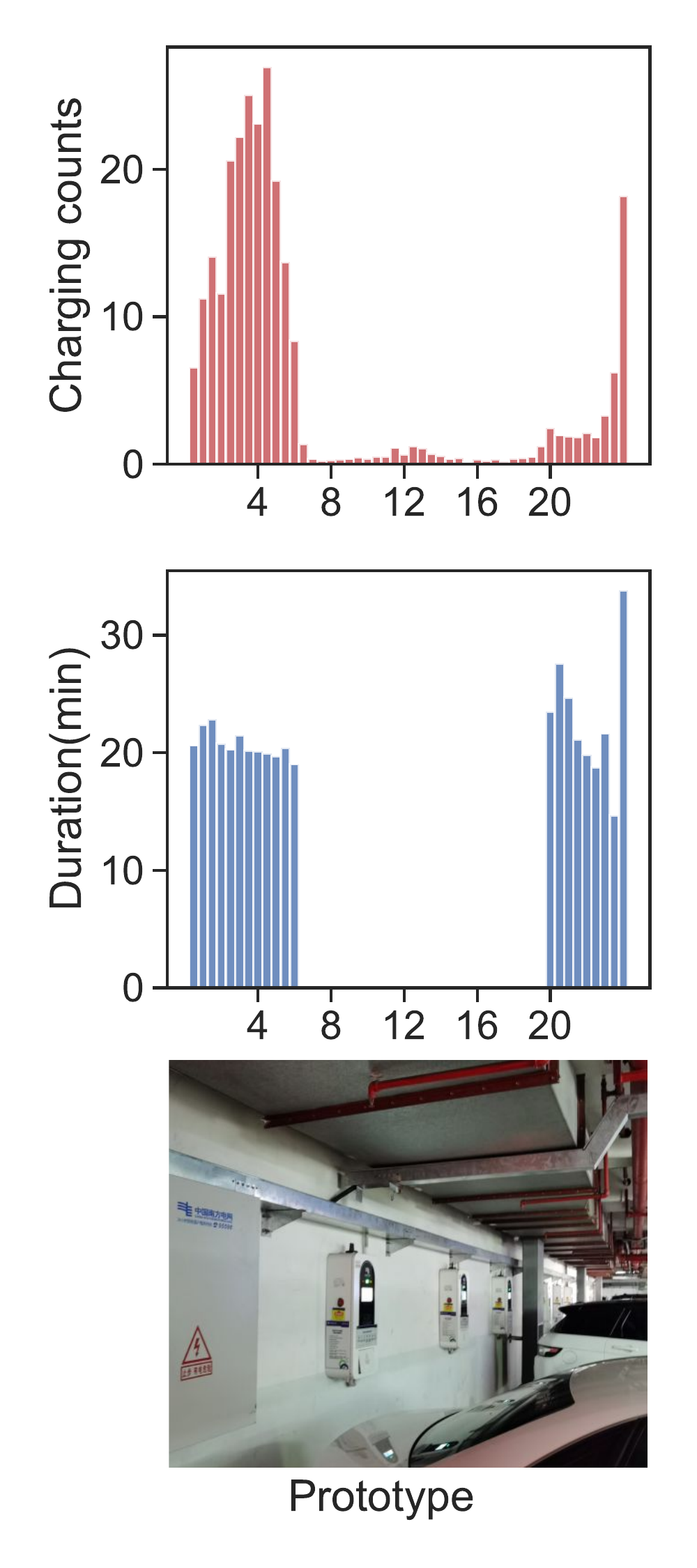}\label{fig:cs_ShoppingPark}}
    \caption{Charging dynamics at the charging station level}
\end{figure}

To further understand the choice of charging stations, we visualize the spatial distribution charging frequencies of all 425 charging stations in Figure~\ref{fig:spatial_cs}. In particular, we select three typical charging stations with distinct daily usage patterns, e.g., Minle P+R station, Zhuzilin station, and Shoping Park station. And their within-day charging dynamics in half-hourly time intervals is shown in Figures~\ref{fig:cs_minle}--\ref{fig:cs_ShoppingPark}. From the spatial distribution, we report that stations close to the city centers (places closer to higher passenger demand) are in general more favored over more distant areas. Nevertheless, we also observe that distinct attributes pertaining to each charging station appear to have greater influence on the charging dynamics than merely the spatial proximity to high demand areas. As an example, Minle P+R station is observed to serve nearly \added{3.7}\% of total daily charging demand. The leading reason for the popularity of Minle station is likely due to its cheaper cost. In particular, Minle station adopted a dynamic pricing structure where the electricity price during off-peak hours is reported to be 30\% lower than peak hours~\cite{sohu}. In addition, a discounted service fee is offered to E-shared mobility vehicles. As a consequence, we find that the time periods where Minle station is most visited (Figure~\ref{fig:cs_minle}) are almost perfectly aligned with the published time for off-peak charging price. Besides its cheaper cost, Minle P+R station is also the largest fast charging station for EVs with 637 fast charging piles. The station is located near the transportation hub and is adjacent to urban expressways, and the station offers rest area with lounge facilities to accommodate the needs of drivers. Moreover, the stations is within walking distance to the Minle metro station. As a consequence, the service reliability, accessibility as well as the level of convenience are likely other factors that make it the top charging station for drivers to take the shift.

\added{Next, we show the usage pattern at the Zhuzilin station (see Figure~\ref{fig:cs_zhuzilin}), which accounts for \added{3.0}\% of total daily charging demand. Similar to Minle P+R station, Zhuzilin also serves as a dedicated charging station for ETs and the station is adjacent to a number of restaurants with affordable price. Compared with Minle, similar usage patterns during noontime peak can be found for its convenience and accessibility. However, one notable difference at Zhuzilin station is during the late-afternoon/early-evening periods (5 PM for Minle and 8 PM for Zhuzilin). We report that one potential reason for the lay-back charging peak is to get prepared for the upcoming demand peak during the early-night period (7-9 PM). In this case, the ETs may charge for longer duration (over 50 minutes during 4-5 PM) or get a quick implement right before the evening demand peak (less than 35 minutes after 7 PM). Further, this location lies in the residential and business area of Futian District with a high demand level. As a consequence, this charging station is more favored during the short lunch break and for an increase opportunity to pick up the next passenger.}

\added{Finally, we present the usage pattern at the Shopping Park station (shown in Figure~\ref{fig:cs_ShoppingPark}). To avoid misleading due to outliers, we only present the average duration in the 30-minute interval with more than one charging count per day. The Shopping Park station is equipped with more than 140 reported charging piles and serves nearly 1.8\% of total charging demand. This station is located in the urban core of Futian District and next to Ping An Finance Center, which is surrounded by bars and restaurants, movie theaters, shopping malls, as well as workplaces. Different from two other dedicated stations, the charging counts during the mid-late night is much higher than other times of the day, which is potentially due to the high demand during the nighttime back to the residential locations. Also, we report the charging peak hours are well aligned with the period of valley electricity price (37.7\% lower than the price in the normal period). Therefore, the high capacity of the charging station, lower electricity price, and increased opportunity for potential high demand at the urban core are three contributing factors that attract the charging demand.}

\begin{figure}[H]
    \includegraphics[width=0.6\linewidth]{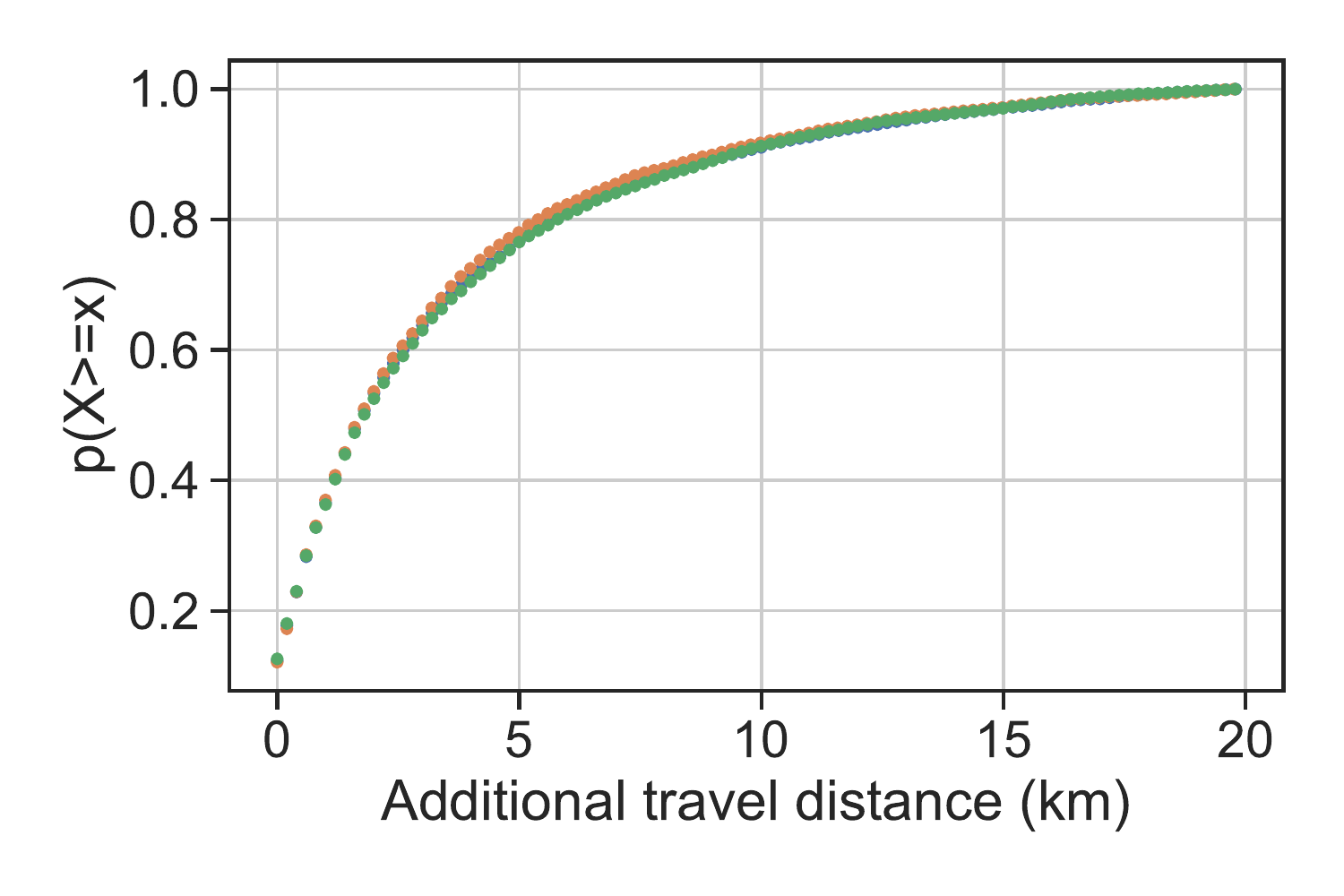}
    \caption{Cumulative distribution of additional travel distance to the charging location as compared to visiting the nearest charging station. }
    \label{fig:additional_travel_distance}
\end{figure}

The charging dynamics at the three selected charging stations highlight the importance of location specific characteristics in attracting charging demand. On the other hand, as a commonly seen and seemingly rational assumption, the distance to the nearest charging station is believed to be another determining factor that drivers the choice of charging locations. We next quantitatively assess how charging distance may be related to the charging behavior, and the results are presented in Figure~\ref{fig:additional_travel_distance}. Here we define additional travel distance as the gap between travel distance from the last dropoff location to the actual charging station and the distance from last dropoff location to the nearest charging station. And we simultaneously visualize the cumulative distribution of 7 days in a week. The important takeaway, as indicated from the figure, is that few ETs visited the nearest charging station. Around 40\% of the drivers will travel up to 1.5 additional kilometers, and 80\% of them may cruise up to 5 extra kilometers for charging. And this observation is consistent across the week. Moreover, there are a small portion of drivers who will travel over 15 kilometers to visit their preferred charging stations. We believe this is most likely due to their last dropoff location is far from the scheduled shift location. \added{These observations are well aligned with the power-law distribution of charging frequency at station level, which indicates that there are some frequently used charging stations which are more preferred by ET drivers. That is, ET drivers may drive for longer distance to charge at preferred charging stations.} As a result, we conclude that ET drivers are less sensitive to the distance to charging station, especially if the distance is below a certain threshold such as 5 kilometers. 

\subsection{Charging pattern analyses}

We use the LDA model~\cite{blei2003latent} to capture the ET driver's charging pattern and temporal regularity. We first manually divide the charging events into single-shift events and double-shift events based on the historical taxi schedule. Furthermore, we select 2,200 single-shift ETs and 7,000 double-shift ETs that appear more than 10 days of the 19 selected weekdays ( a Friday, 9/13/2019, is neglected due to Mid-Autumn Festival). To determine the optimal LDA model, we next partition the ETs into training and validation sets (1,800 vs. 400 for single-shift ETs and 6,300 vs. 700 for double-shift ETs). Finally, we conduct a grid search procedure with \added{the spatial dimension set, $J=\{10,20,30,40,50\}$ and the temporal dimension set $K = \{6,8,10,12,14,16\}$}. Based on the perplexity scores, the combination of \added{$(J=30,K=10)$ and $(J=30,K=12)$} are determined as the optimal grid for the single-shift and double-shift ETs, respectively. Other  hyperparameters ($\alpha=\beta=\gamma=0.01$) in the LDA model are consistent with Sun et al.~\cite{sun2021routine}. 

\subsubsection{Individual-level charging activity pattern}

\begin{figure}[H]
    \centering
    \subfloat[Single-shift ETs]{\includegraphics[width=\linewidth]{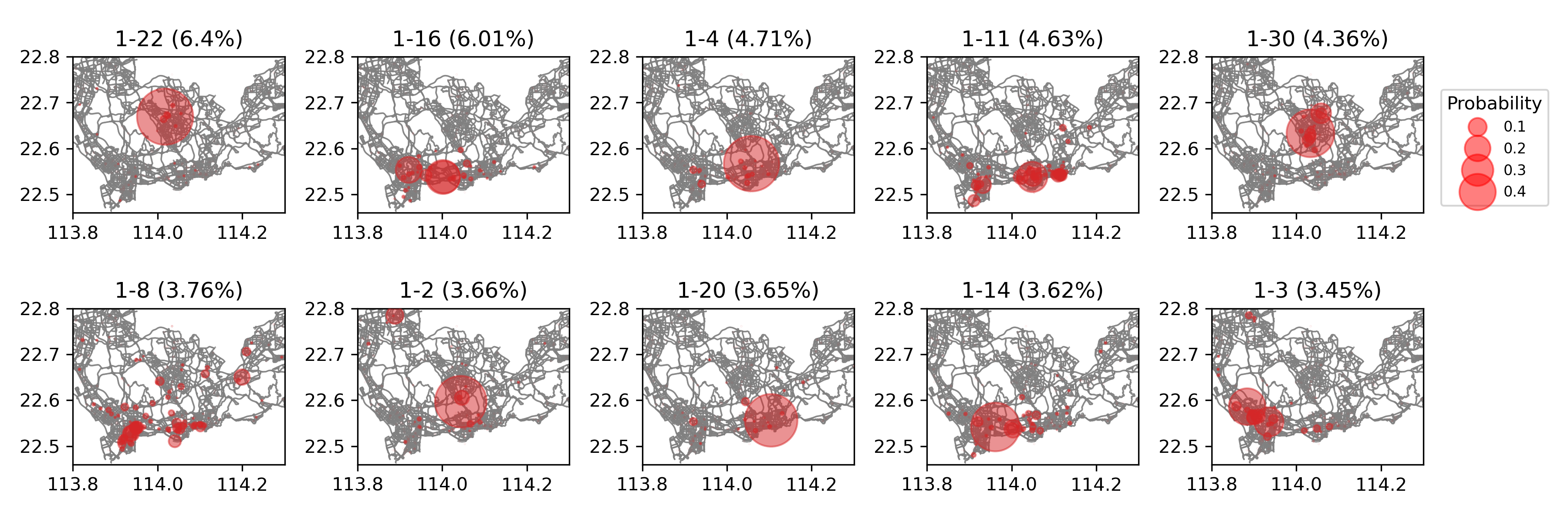}\label{fig:single_spatial_pattern}}\\
    \subfloat[Double-shift ETs]{\includegraphics[width=\linewidth]{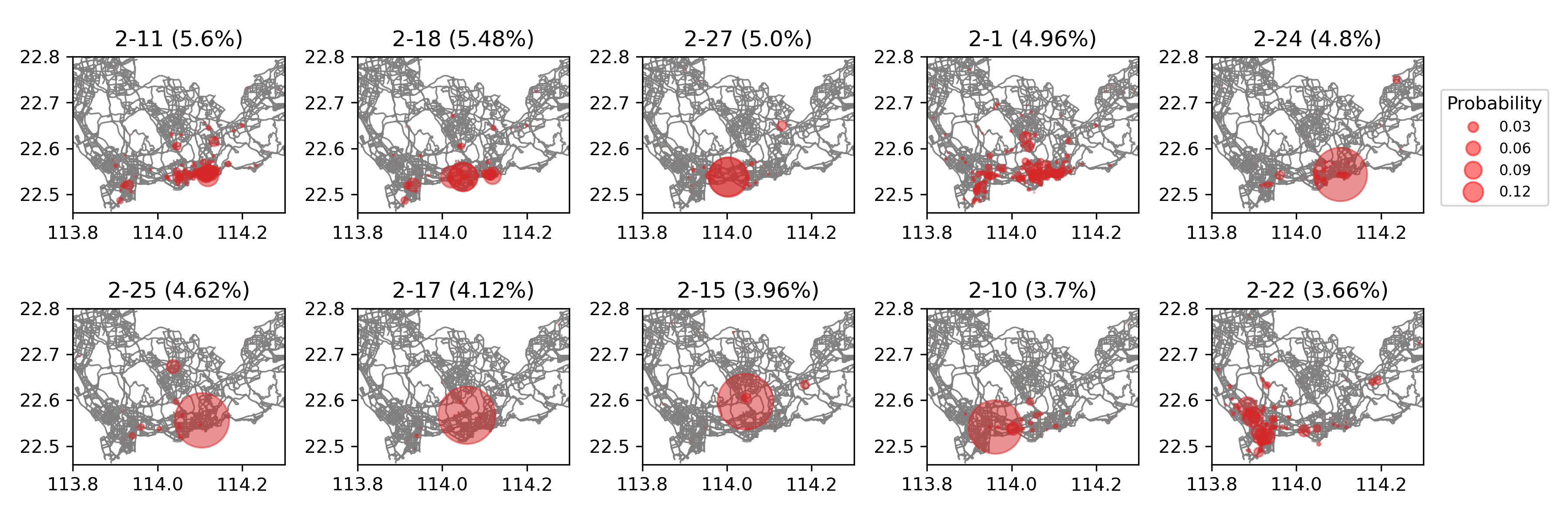}\label{fig:double_spatial_pattern}}
    \caption{Selected spatial patterns of charging activities}
    \label{fig:spatial_charging_patterns}
\end{figure}

With the optimal pattern numbers $J$ and $K$, we present the spatial and temporal charging patterns that are summarized from the charging events per ET over four weeks. Here we present 10 most dominant spatial patterns both the single- and double-shift ETs in Figure~\ref{fig:spatial_charging_patterns}, where the corresponding percentages are attached in the brackets. The overview of all spatial patterns is presented in~\ref{sec:all_spatial_patterns}. In the figures, the size of each circle represents the probability of a charging event at the specific charging stations. Based on the results, we summarize three major spatial charging patterns as follows:

\begin{itemize}
    \item \textbf{Planned charging in suburbs and peripheries }(e.g., Patterns 1-22, 1-30, 1-2, and 2-15): this feature shows the charging activities at several dedicated charging stations (e.g., Minle P+R station) located in the suburbs and peripheries, which may indicate less waiting time but with sufficient supplies of charging piles.
    \item \textbf{Planned charging in city centers }(e.g., Patterns 1-4, 1-20, 1-14, 2-27, 2-25, 2-17, and 2-10): these patterns denote the common choices when single-/double-shift ETs need to charge. In the planning stage, more charging piles are required so as to accommodate the extensive charging demand in the city centers.
    \item \textbf{Unplanned charging in city centers }(e.g., Patterns 1-16, 1-11, 1-8, 1-3, 2-11, 2-18, 2-1, 2-22): in this feature, the corresponding charging stations may encounter the local congestion in some time periods such that the ETs will detour to find the available charging piles nearby.
\end{itemize}

By considering all spatial patterns (see Figures~\ref{fig:all_single_spatial_pattern} and~\ref{fig:all_double_spatial_pattern}), we report that the proportion of spatial patterns in city centers, suburbs, and peripheries are 61.84\%, 16.82\%, and 21.30\% for single-shift ETs and, 69.81\%, 23.26\%, and 6.89\% for double-shift ETs. In this regard, compared with double-shift ETs, more single-shift ETs tend to charge in the peripheries. This may be because the single-shift ETs have a more flexible time schedule such that the ET drivers have sufficient time to rest at those peripheral charging stations equipped with accessible lounges or restaurants. On the other hand, we also observe a significantly larger proportion of double-shift ETs that charges around the city centers ($69.81\%$ versus $61.84\%$). This indicates that the double-shift ET drivers tend to make efficient usage of their on-service time, which in turn translates to limited time for the charging-related activities (waiting, charging, and traveling to and from charging stations), resulting in more charging activities in city centers.

\begin{figure}[H]
    \subfloat[Single-shift ETs]{\includegraphics[width=\linewidth]{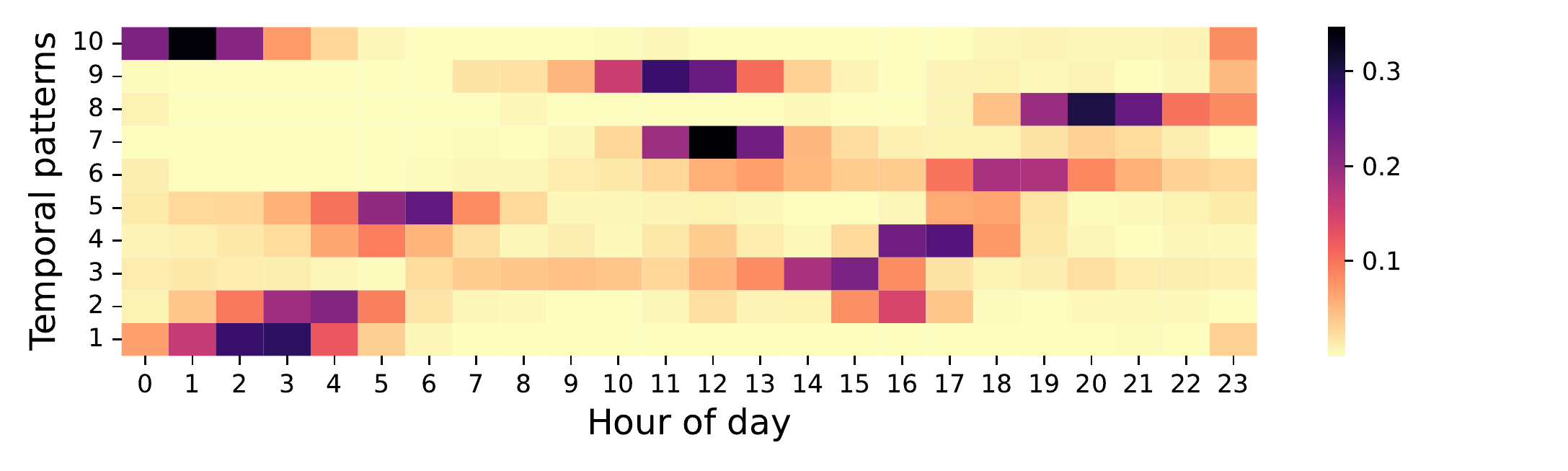}\label{fig:single_t_pattern}}\\
    \subfloat[Double-shift ETs]{\includegraphics[width=\linewidth]{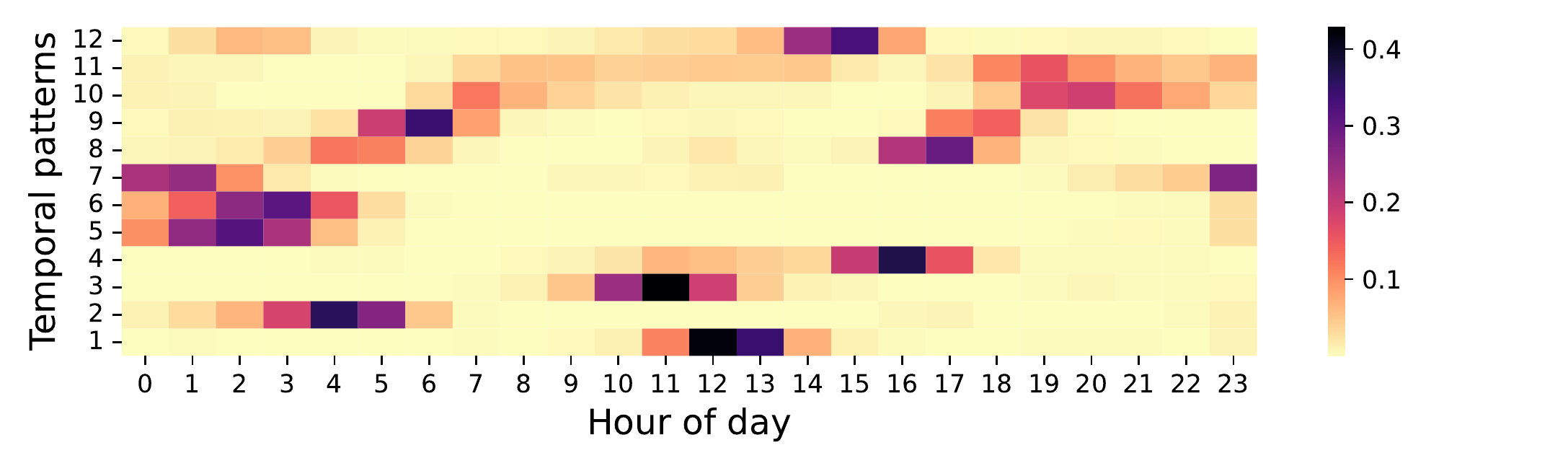}\label{fig:double_t_pattern}}
    \caption{Temporal patterns of charging activities}
    \label{fig:temporal_charging_patterns}
\end{figure}

Aside from the spatial pattern, we also explore the temporal patterns for the single- and double-shift ETs. Three peaks of charging activities can be observed regarding the late night/early morning (1-6 AM), around noon (11 AM - 2 PM), and afternoon/early evening (4 - 6 PM). For the single-shift ETs, three particular patterns (Patterns 1,2,10) indicate the charging events at late-night periods (12 AM - 4 AM), which may capture the times when ET drivers rest and conduct \added{charging activities near home locations}. In addition, two-mode temporal patterns can be found in single-shift ETs (e.g., see Figure~\ref{fig:single_t_pattern} Pattern 2, 3, and 5), which can be interpreted as the charging before-and-after the daily operation. As for the double-shift ETs, the ET drivers prefer to charge before the morning and evening peaks (about 6 AM and 5 PM), so as to provide mobility service with less disturbance for charging and gain more revenue from the extensive demand. Similarly, the two-mode temporal patterns (e.g., Patterns 8, 9, 10, 11, and 12) are also found in the double-shift ETs, which suggests the potential shift time between two ET drivers. For such patterns, we report that the shift time does not strictly follow the morning-evening shifts, but the time may slightly deviate, e.g., 5 AM - 5 PM, 6 AM - 6 PM, and 7 AM - 8 PM as shown in Pattern 8, 9, and 10 in Figure~\ref{fig:double_t_pattern}, respectively. Finally, we note that the around-noon activities are observed in both ET patterns (at around 11 AM - 2 PM), implying that the ET drivers may have lunch or take a break while charging.

By considering both spatial and temporal dimensions, we present the spatiotemporal charging pattern with the corresponding probability distributions on each axis. From the aspect of temporal dimension (the vertical bars), the patterns with high probabilities mainly capture the late-night/early-morning charging during 12 AM - 3 AM for single-shift ETs (see Pattern 1 and 10 in Figure~\ref{fig:single_t_pattern}) and during 1-4 AM double-shift ETs (see Pattern 5 and 6 in Figure~\ref{fig:double_t_pattern}). 
As for the single-shift ETs on the spatial dimension, two significant patterns are observed with high probabilities (Patterns 1-16 and 1-22 with $6.4\%$ and $6.01\%$), indicating the planned charging patterns in the periphery, which evidences the relatively sufficient time for the single-shift ET drivers. On the other hand, the frequent patterns of double-shift ETs are more likely to be charging with preference in the city centers (Patterns 2-11 and 2-18 with $5.6\%$ and $5.48\%$). In this case, the ET drivers can ensure high service efficiency with minor disturbance due to charging.

\begin{figure}[H]
    \subfloat[Single-shift ETs]{\includegraphics[width=0.5\linewidth]{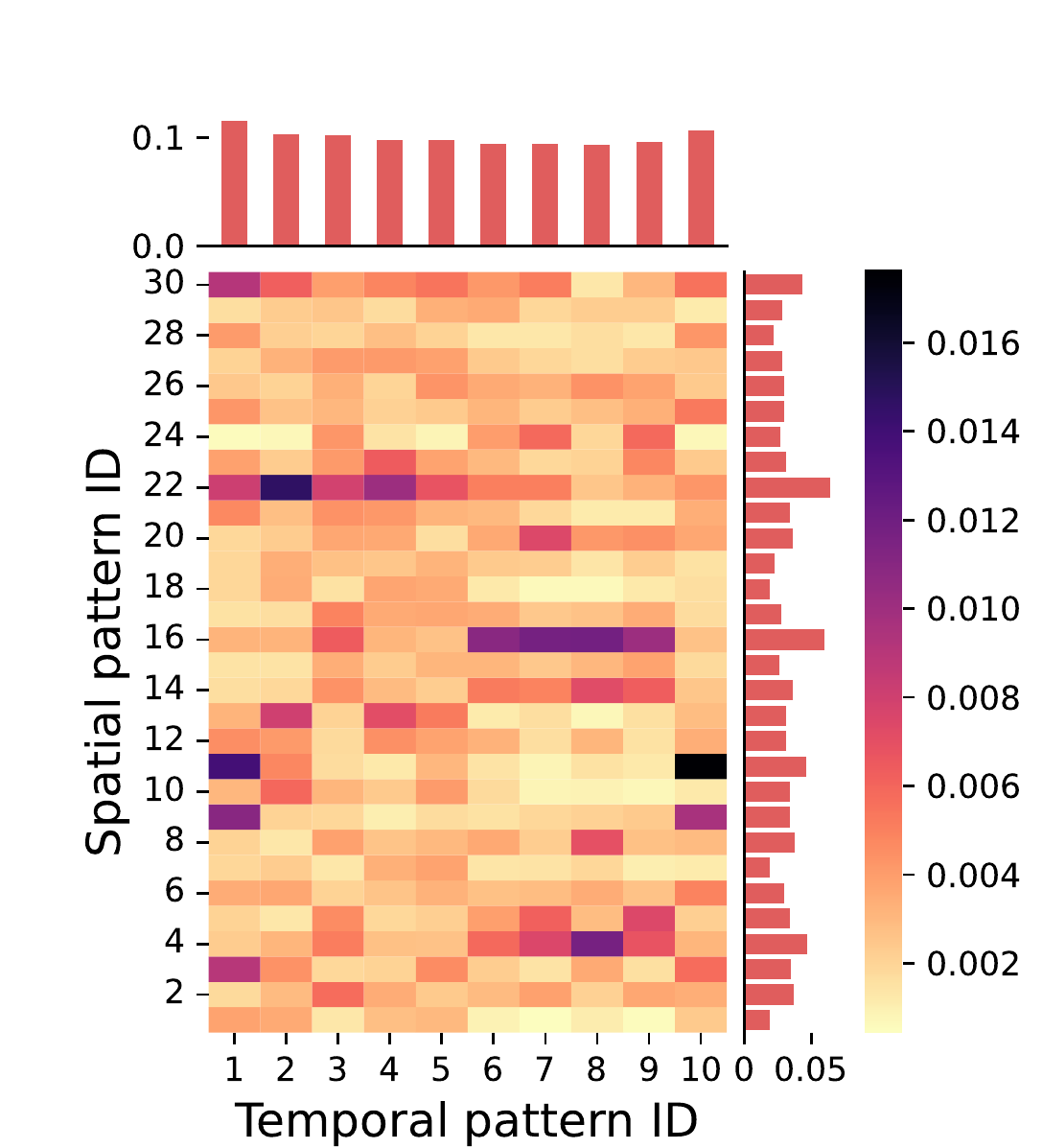}\label{fig:single_st_pattern}}
    \subfloat[Double-shift ETs]{\includegraphics[width=0.5\linewidth]{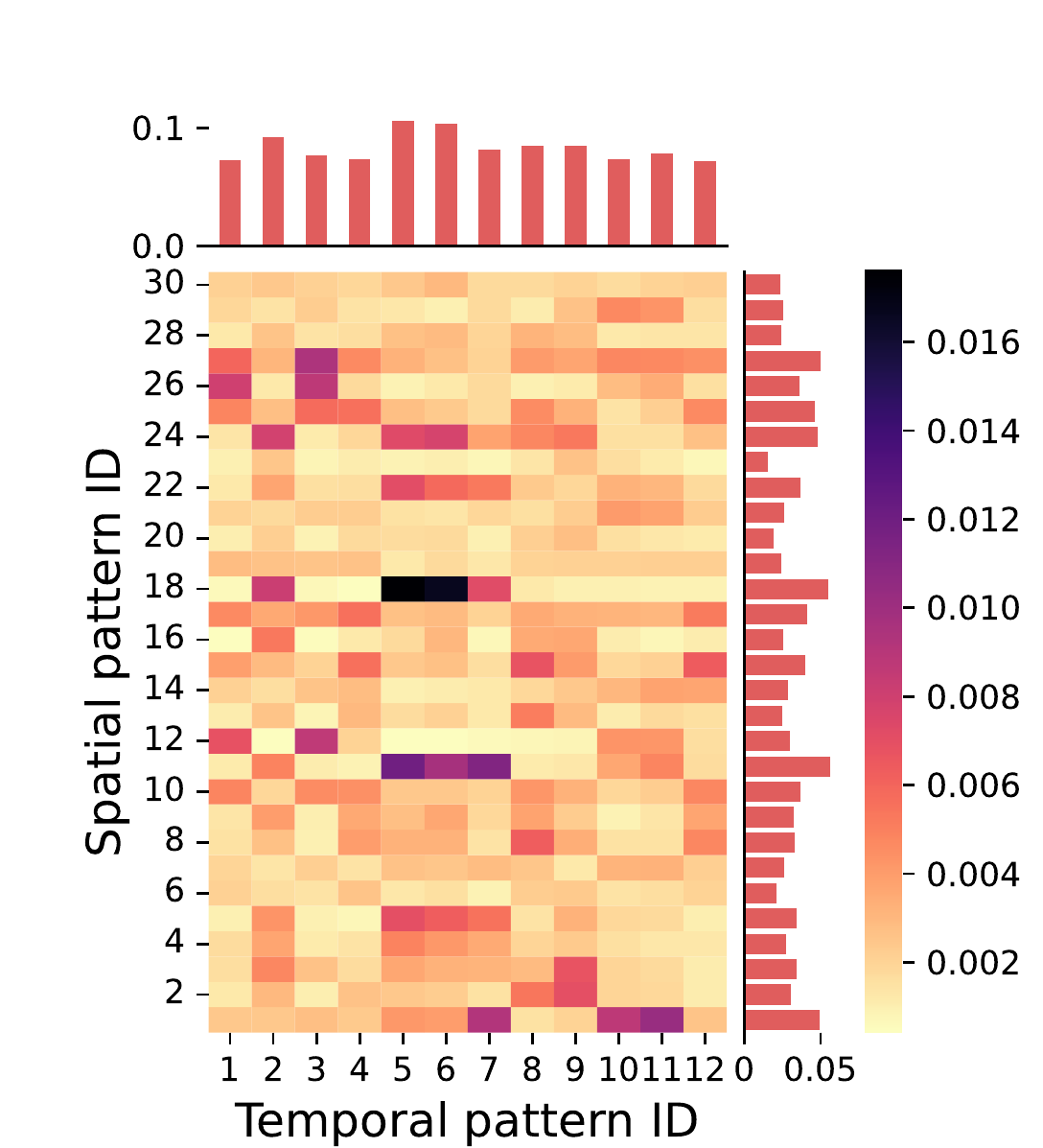}\label{fig:double_st_pattern}}
    \caption{Spatiotemporal patterns of charging activities}
    \label{fig:spatiotemporal_charging_patterns}
\end{figure}

Furthermore, we summarize several spatiotemporal patterns with high probability, as seen in the dark-colored cells in Figure~\ref{fig:single_st_pattern}. 
We report that one major pattern is the cell $(2,22)$, which denotes the double-mode charging in the peripheral area. With a further look in the pattern ``1-22", the dominant charging station is within a urban village in Longhua District, which is surrounded by several populated residential areas and one transportation hub. As such, it is of great potential to attract the ET drivers who live nearby to charge before and after the daily operation. 
Also, given the spatial pattern of ``1-11", two temporal patterns suggest the planned charging in city centers during the late-night period (the cell $(1,11)$ and $(10,11)$). The charging stations in city centers are potentially near the bars and restaurants, and also provides the time-of-use electricity price, which is particularly attractive during the late-night period. One example is the charging activities in the Shopping Park station.
Similar physical interpretation can also be captured in the double-shift ETs regarding the three significant spatiotemporal cells: $(5,18), (6,18), (7,18)$, denoting the planned charging in city centers during the late-night/early-morning period. It implies that the double-shift ET drivers have to follow the shift schedule spanning about 12 hours and prefer to charge near the high-demand locations. 
As such, we conclude that those common features can well reveal the latent spatiotemporal patterns that underline in both single- and double-shift ETs.

\subsubsection{Regularity of individual charging activities}

After presenting the charging patterns in spatial and temporal dimensions, we finally investigate the underlying spatiotemporal regularity of the individual charging behavior by computing the average cosine similarity over one ET's weekly charging event. Specifically, we only select the 19 weekdays of the first four weeks in September 2019, where the Mid-Autumn Day (a Friday) is omitted. Next, for each ET, we generate sequences of charging events in a rolling time window fashion with a length of 5. In this regard, we treat the charging events covered by one rolling window for each ET as a document. And the output is a collection of vectors, where each vector represents the probabilities of one sequence of charging events being assigned to each spatiotemporal pattern. Finally, the spatiotemporal regularity of each ET is quantified by the average cosine similarity score, measured by the cosine distance between all pairwised vectors associated with the particular ET. Intuitively, a high similarity score indicates a high level of regularity such that future charging patterns can be accurately predicted based on historical records. Note that we use the same set of ETs in training set for the charging pattern analyses in the previous section (1,800 for single-shift ETs and 6,300 for double-shift ETs).

\begin{figure}[H]
    \subfloat[Single-shift (weekly) \label{fig:regularity_single_week}]{\includegraphics[width=0.45\linewidth]{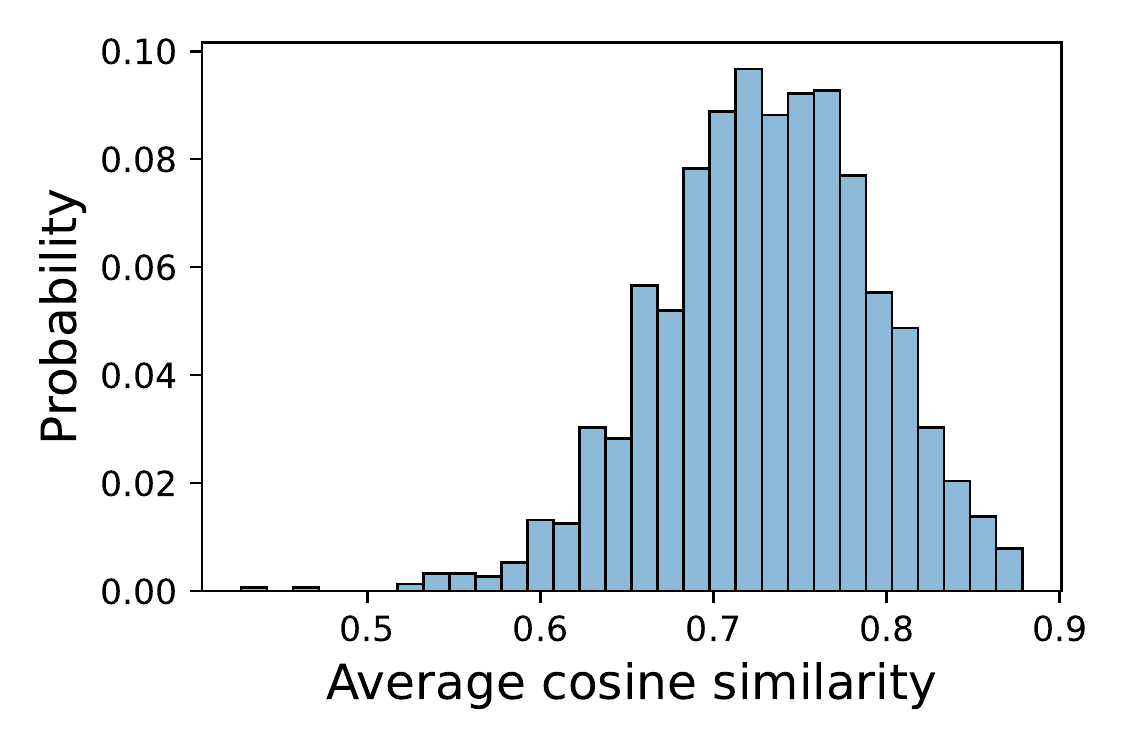}}
    \subfloat[Double-shift (weekly)\label{fig:regularity_double_week}]{\includegraphics[width=0.45\linewidth]{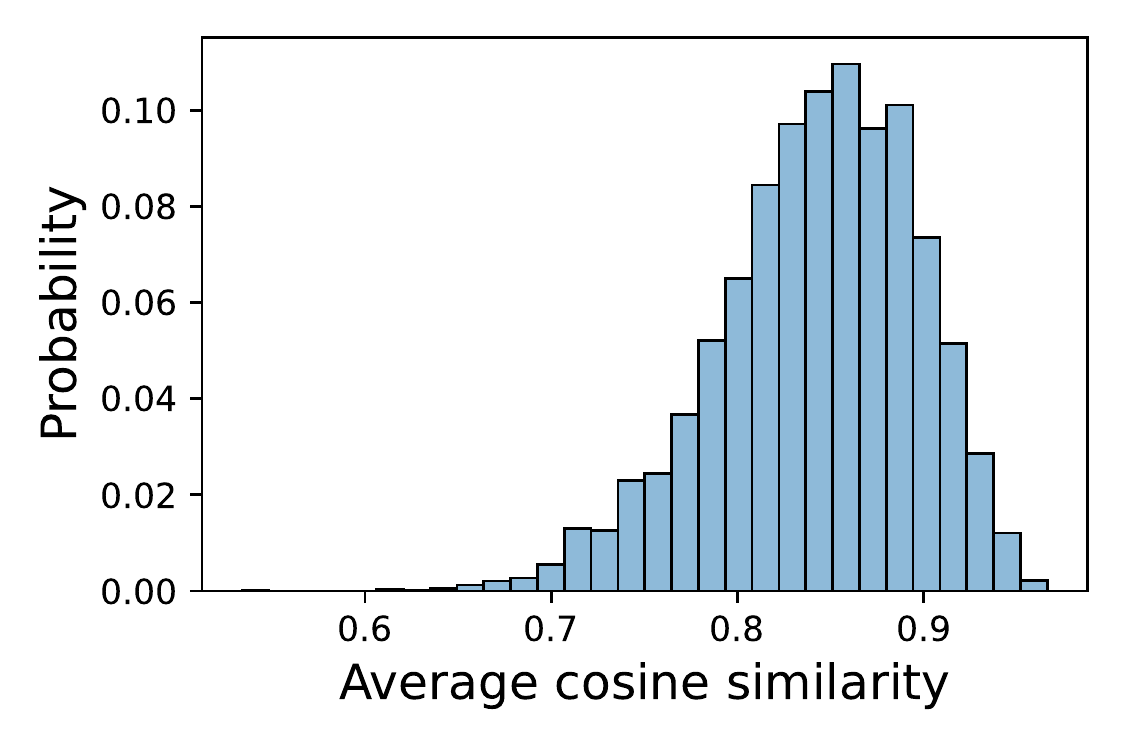}}
    \caption{Dynamics of individual temporal regularity.}
    \label{fig:individual_similarity}
\end{figure}

As seen in Figures~\ref{fig:regularity_single_week} and~\ref{fig:regularity_double_week}, we report a slightly left-skewed distribution, indicating that most of the ETs conduct charging activity following similar charging patterns on the spatiotemporal dimensions. 
Specifically, the average similarity scores of double-shift ETs are higher than the single-shift ETs (0.84 vs. 0.73). The higher level of similarity may result from the more generalized shift time along the day as observed in the temporal patterns for double-shift ETs. In addition, a few observations with a score under $0.6$ are also found in single-shift ETs. It suggests the anomaly level over each week, implying more flexibility for the single-shift ET drivers. 
Finally, we highlight that these similarities in the spatiotemporal charging pattern will shed light on charging stations' coordinated management to satisfy the charging demand at similar time periods and specific spots. More detailed discussion and implications are presented in the next section.



\section{Discussion}
Based on the results at the system level, charging station level and individual driver level, we summarize the following five major findings and discuss their implications for charging planning and management of ETs, and potentially for E-shared mobility systems in general. 




\textbf{Finding 1: The system-level charging dynamics show spatiotemporally unbalanced intraday charging station usage, which is observed to be stationary across days.}

From the aspects of the ET operation and management, the spatiotemporally unbalanced distribution of charging activities leads to the prolonged waiting time at the charging stations during peak hours, and subsequently, a shortage of supply for mobility services. While the charging behavior of ETs is closely related to their shift schedules, policies and guidelines may be necessary to motivate a staggered shift schedule for ETs with two shifts.  This finding also has significant implications for the modeling and planning of charging facilities. The stationary dynamics support the appropriateness of the assumption on equilibrium behavior for system models. However, caution should be exercised in terms of the assumed utility function that drives the equilibrium solutions. 

\textbf{Finding 2: Preferential attachment behavior for usage at the charging station level.}

A considerable number of charging facilities are expected to be developed to cope with the booming of EVs worldwide. However, given a budget of investment, it is important to consider the realization of a skewed facility usage pattern at each charging station. It is especially so with the construction of charging hubs, which brings in the scale of the economy (more charging piles with reduced charging cost and easy access to other facilities), which can offer better personal utility for ET drivers. In this regard, simply increasing spatial coverage of the charging station may not be necessary and will likely result in a waste of resources. Moreover, special attention should be paid to the coupling between the mobility service system and the reliability of the power system. The power-law charging visits distribution implies a charging infrastructure system that is efficient yet vulnerable. The concentration of charging needs may result in the local failure of the power grid due to demand surge, which will backpropagate and cause cascading failure in the mobility service systems. \added{Therefore, cautious is much needed} to develop a failure-resilience charging infrastructure plan and backup operation strategy at major charging hubs in case of a facility failure.

\textbf{Finding 3: Stationary dynamics at system-level is the result of strong charging activity regularity at the individual driver level.}

While system-level stationarity may be achieved via frequent but symmetric changing behavior from individual drivers, we report that the individual-level behavior shows very strong regularity over time for the majority of the drivers. However, this strong regularity is a mixture of unplanned charging in the city center and planned to charge activities in other regions. This shapes a sharp contrast to the refueling behavior of gasoline vehicles. The finding also suggests the possibility for accurately predicting exact charging activities at many of the charging stations. As such, modeling approaches based on individual charging activity sequences can be established and will contribute to much more efficient operational management of the rate of charge that may best serve EVs for mobility services.

\textbf{Finding 4: Factors other than the distance to the charging station have a greater influence on drivers' preference for the location of charging stations.}

This finding echos our discussion on the choice of utility function for driving equilibrium behavior. It is noted that many existing studies considered spatial proximity as the most influential factor, which we report may barely hold for the modeling of E-shared mobility services. Instead of visiting the nearest charging station, most drivers cruise extra kilometers to a charging station that best aligns with their other interests, including but not limited to the capacity of a charging station, the charging cost, the convenience for making a shift, the availability of supporting facilities such as the rest area and restaurants, and even their acquaintance of a charging station. As this outcome contradicts the assumption of many existing studies, it highlights the pressing need to quantitatively assess the determining factors that best characterize the utility of a charging activity for heterogeneous drivers of E-shared mobility services. 

\textbf{Finding 5: Overnight charging is not viable for most of the drivers, especially for double-shift ET drivers.}
  
Another common assumption, if not a misperception, is that E-shared mobility drivers may charge overnight (e.g., at home), and charging activities during the daytime are less important. Through our analyses, we argue that overnight charging is not viable for the majority of the drivers for E-shared mobility services. On the one hand, overnight charging implies owning a private parking spot equipped with charging piles. This is, however, unlikely for E-shared mobility drivers given the excessively high cost. On the other hand, even at charging facilities, we report that only 35.8\% of all charging activities occur between 12 AM and 7 AM. Out of these activities, 0.86\% of the charging activities have a duration longer than 4 hours. This implies that drivers are also unlikely to charge overnight at the charging stations.  \added{It should be noted that although the present work only identifies charging activities at public charging stations, it has been examined by the ground truth survey that most of the ET drivers (over 95\%) use public charging stations other than private ones, which could further support our observations regarding overnight charging.} One primary reason is to avoid the overtime cost (e.g., parking cost at charging stations). Consequently, the coordination of daytime charging for E-shared mobility services becomes an important research topic that requires extra attention. Moreover, it may be worthwhile considering the introduction of dedicated park-and-charge facilities, which may best prevent the overflow of charging demand to the power grid during the daytime. Doing so may lead to a win-win situation that reduces the E-mobility system's charging cost and improves the power system's reliability.

\section{Conclusion}


In this study, we investigated the spatial-temporal charging dynamics of ETs on the system level and also explored the individual-level ET driver's charging behavior and regularities using ETs' GPS trajectory data in Shenzhen for four weeks in 2019. From the temporal distribution, we report three charging peaks over one day during late-night/early-morning periods, noontime, and late-afternoon/early-evening periods. On the other hand, the usage pattern of the charging station is observed to follow the power-law distribution, which necessitates more tailored decision-making on the planning stage so as to make full usage of the charging resources. Furthermore, the spatial and temporal correlation analyses suggest strong intraday regularities and interday similarities. Besides, we capture the charging patterns on both system and individual levels via the two-dimensional LDA model. The results suggest that most of the charging events can be categorized based on several key features, e.g., charging peaks and preference on charging stations for both single- and double-shift ETs. Finally, we quantitatively analyze the regularity of charging patterns on an individual level using the cosine similarity. Most ETs exhibit a similarity score of over 0.8, indicating a high level of regularity in terms of individual charging preferences, which is a driving force to the system-level stable charging dynamics. 

We note that the identified charging events serve as a subset of all charging activities due to the data limitation on the specified charging status, the latest layout of the charging stations, and the level of accuracy of the GPS trajectory data. But a large number of identified charging activities are sufficient to lay out statistically meaningful results that best capture the charging dynamics of a city-scale E-shared mobility system. For future studies, efforts can be made towards dedicated charging station planning based on the key features of the ET drivers' interests and preferences. Meanwhile, a charging station recommendation algorithm incorporated by pricing strategies will help to (1) balance the skewed usage pattern of the charging resources and (2) coordinate the charging activities to minimize the waiting time at charging stations. \added{In addition, as the study depicted the initial dynamics on charging behavior of ET drivers, it is important to perform a comprehensive assessment by modeling the influencing factors and understanding how they affect the choice of charging stations.}
\added{Finally, under a collaboration with Shenzhen Transportation Planning Center, we can only access the one-month trajectory data. A more comprehensive study, e.g., long-term evolution of ET charging patterns, can be conducted if the data is available to support the investigation over a longer time horizon.}

\section{Acknowledgments}
This research was partly supported by Research on Trip Chain of Public Transit Passenger by 
Research on the Coordination Strategies and Methods of Transport Supplies to Satisfy the Tempo-spatial Pattern of Public Transit Transferring in the Context of Data Fusion from Multiply Sources (General Research Project of Humanities and Social Sciences of the Ministry of Education under Grant No.21YJC630029), the Guangdong Basic and Applied Basic Research Foundation (No. 2020A1515111001),
Research on Guidance Management of Travel Demand in Metropolitan Bay Area based on Travel Pattern (Philosophy and Social Science Planning Project of Guangdong Province under Grand No. GD20CGL30) and Development and Application of Simulation System of Travel Behavior based on Data Fusion in the Context of V2X (Natural Science Foundation of Top Talent of SZTU under Grant No. 20200218). 

\bibliographystyle{unsrt}
\bibliography{ref}

\newpage
\appendix

\section{Complete results of spatial charging patterns}~\label{sec:all_spatial_patterns}


\begin{figure}[H]
    \includegraphics[width=\linewidth]{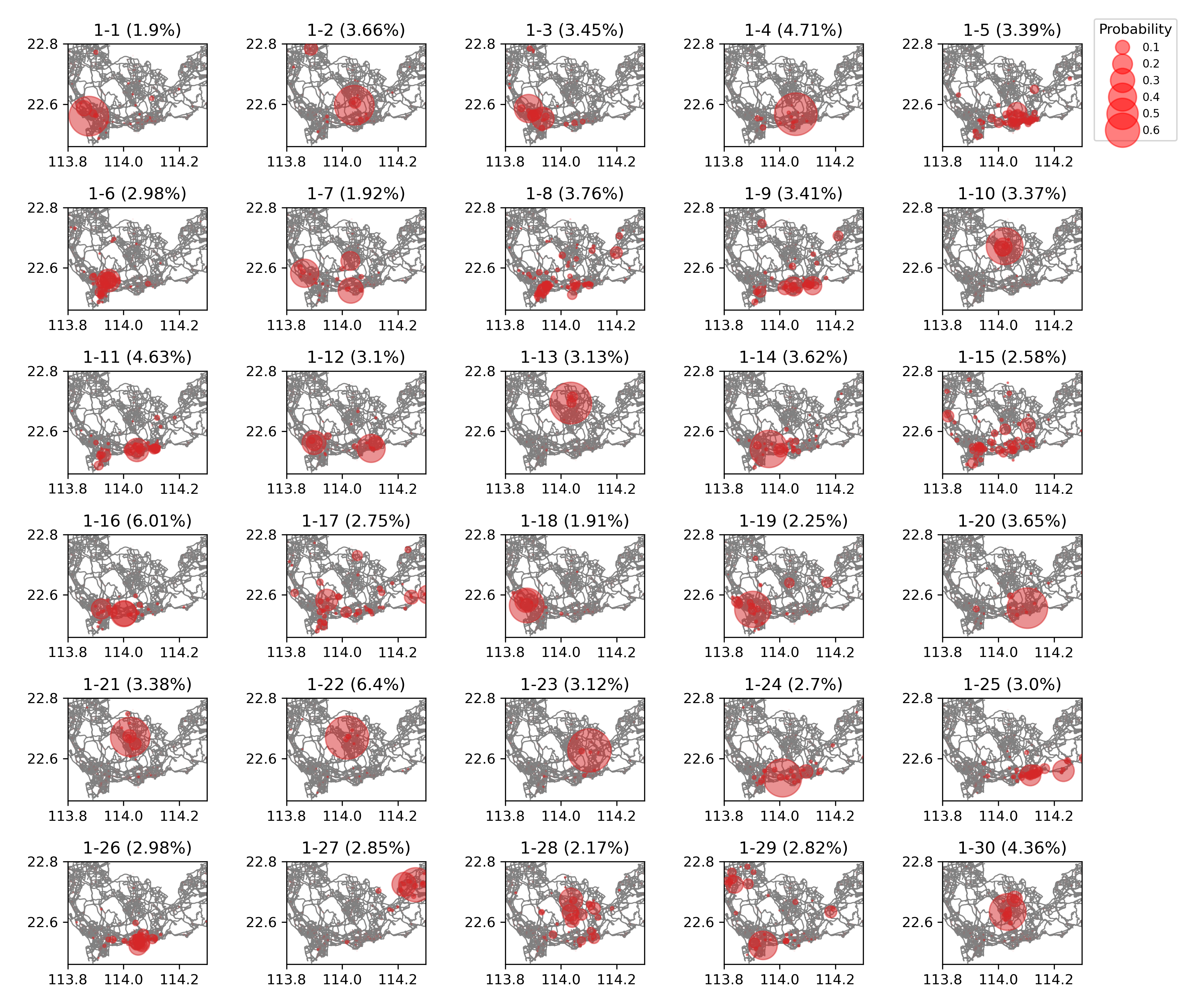}\caption{Charging activities of single-shift ETs}
    \label{fig:all_single_spatial_pattern}
\end{figure}

\begin{figure}[H]
    \includegraphics[width=\linewidth]{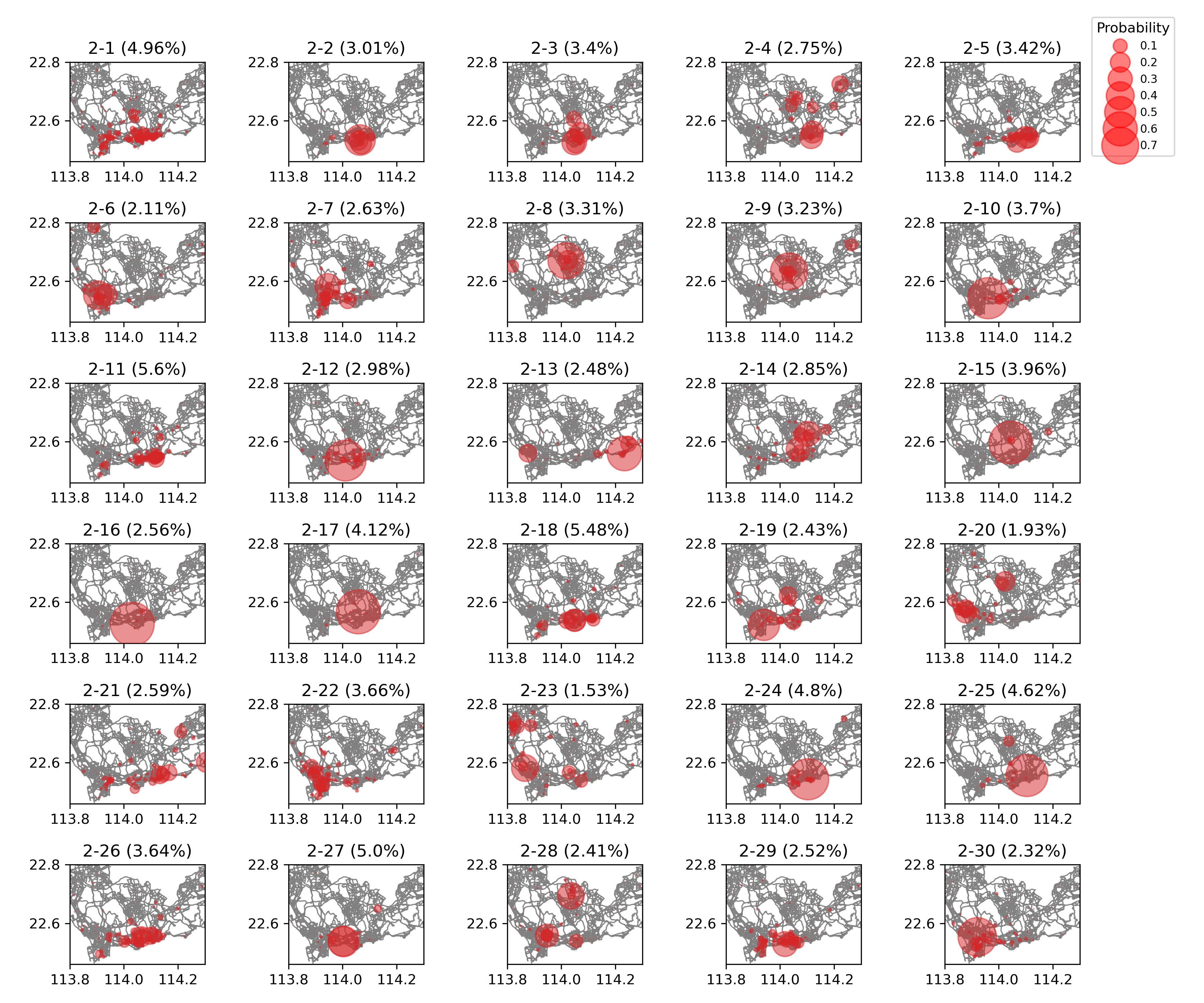}\caption{Charging activities of double-shift ETs}
    \label{fig:all_double_spatial_pattern}
\end{figure}

\newpage
\section{Questionnaire for Shenzhen ET drivers }~\label{sec:questionnaire}

\begin{figure}[H]
    \includegraphics[width=.9\linewidth]{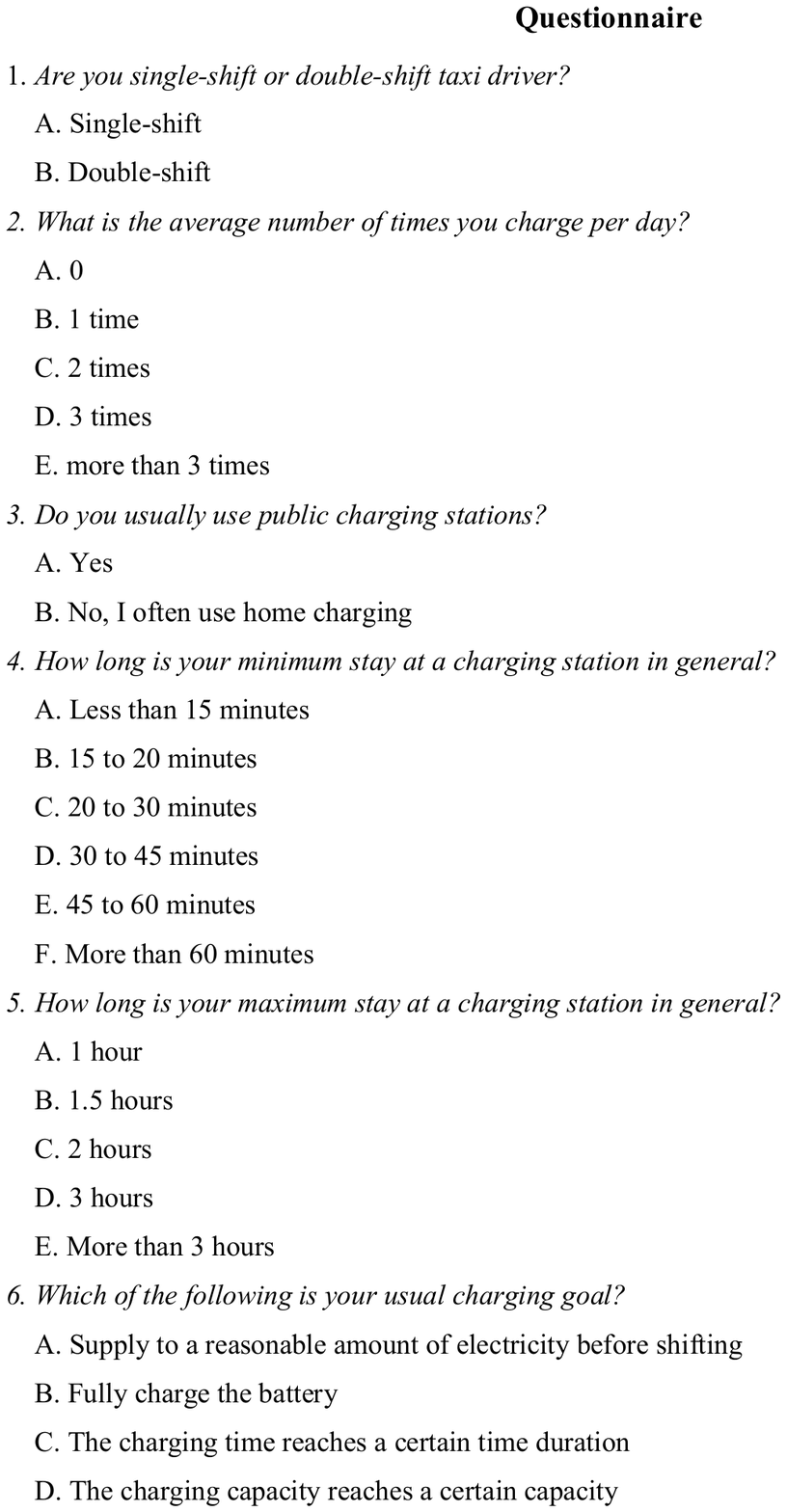}
    \label{fig:question}
\end{figure}

\end{document}